\documentclass[12pt,preprint]{aastex}

\newcommand{\etal }{{et al.} }
\newcommand{\msun}{\thinspace M_\odot}

\newcommand{\vect}[1]{\mbox{\boldmath$#1$}}
\def\lesssim{\mathrel{\hbox{\rlap{\hbox{\lower4pt\hbox{$\sim$}}}\hbox{$<$}}}}
\def\gtrsim{\mathrel{\hbox{\rlap{\hbox{\lower4pt\hbox{$\sim$}}}\hbox{$>$}}}}
\newcommand{\cm}{\,{\rm cm}^{-3} } 
\newcommand{\km}{\,{\rm km\, s}^{-1}} 
\newcommand{\nc}{n_{\rm c} } 
\newcommand{\rcri}{R_{\rm c} }
\newcommand{\mdot}{M_\odot\,{\rm yr}^{-1} }

\newcommand{\rhoc}{\rho_{\rm c}}
\newcommand{\dfrac}[2]{{\displaystyle \frac{#1}{#2}} }
\newcommand{\tffc}{t_{\rm ff,c}}
\newcommand{\tffb}{t_{\rm ff,b}}

\shorttitle{Protostellar Outflow}
\shortauthors{Machida \& Matsumoto 2010}

\begin{document}

\title{Impact of Protostellar Outflow on Star Formation: Effects of Initial Cloud Mass}
\author{Masahiro N. Machida\altaffilmark{1} and Tomoaki Matsumoto\altaffilmark{2}} 
\altaffiltext{1}{Department of Earth and Planetary Sciences, Faculty of Sciences, Kyushu University, 6-10-1 Hakozaki, Higashi-ku, Fukuoka 812-8581, Japan; machida.masahiro.018@m.kyushu-u.ac.jp}
\altaffiltext{2}{Faculty of Humanity and Environment, Hosei University, Fujimi, Chiyoda-ku, Tokyo 102-8160, Japan; matsu@i.hosei.ac.jp}

\begin{abstract}
Star formation efficiency controlled by the protostellar outflow in a single cloud core is investigated by three-dimensional resistive MHD simulations. 
Starting from the prestellar cloud core, the star formation process is calculated until the end of the main accretion phase.
In the calculations, the mass of the prestellar cloud is parameterized.
%% after going through the protostar formation
During  the star formation, the protostellar outflow is driven by the circumstellar disk.
The outflow extends also in the transverse direction until its width becomes comparable to the initial cloud scale, and thus, the outflow has a wide opening angle of $\gtrsim40\degr$. 
As a result, the protostellar outflow sweeps up a large fraction of the infalling material and ejects it into the interstellar space. 
The outflow can eject at most over half of the host cloud mass, significantly decreasing star formation efficiency.
The outflow power is  stronger in clouds with a greater initial mass.
Thus, the protostellar outflow effectively suppresses star formation efficiency in a massive cloud.
The outflow weakens significantly  and disappears in several free-fall timescales of the initial cloud after the cloud begins to collapse.
The natal prestellar core influences  the lifetime and size of the outflow.
At the end of the main accretion phase, a massive circumstellar disk comparable in mass to the protostar remains.
Calculations show that typically, $\sim30$\% of the initial cloud mass is converted into the protostar and $\sim20$\% remains in the circumstellar disk, while $\sim40$\% is ejected into the interstellar space by the protostellar outflow. 
Therefore,  a single cloud core typically has a star formation efficiency of $30-50$\%.

\end{abstract}

\section{Introduction}
\label{sec:intro}
Many protostellar outflows are observed in star-forming regions.
These outflows are believed to be  universally driven by a protostar and play a critical role in star formation\citep{arce07,bally07}. 
The protostellar outflow may determine star formation efficiency, especially in low-mass star formation process.
The similarity between the core mass function and the initial mass function implies that only a certain fraction of the prestellar core is converted into the star \citep{motte98,andre09}.
Recent observations have shown that, in a single prestellar core, star formation efficiency is limited to $\epsilon \equiv M_{\rm star}/M_{\rm core}\sim20-50\%$ \citep{andre10,konyves10}.
\citet{matzner00} argued that star formation efficiency might be limited by the protostellar outflow. 
With a simple analytical approach, they showed  that a wide-opening-angle outflow sweeps up the gas in the infalling envelope and ejects it into the interstellar space.
They concluded that the protostellar outflow can limit star formation efficiency to $\epsilon \sim30-50\%$.
In addition, feedback from the protostellar outflow may maintain interstellar turbulence and affect subsequent star formation \citep{nakamura07}.
Thus, the protostellar outflow is crucial in (low-mass) star formation.
However, it is quite difficult to model the outflow using only an analytical approach because the outflow driving depends on various conditions such as the infalling mass rate onto the circumstellar disk, the size of the circumstellar disk, the configuration and the strength of the magnetic field around the protostar.
Thus, numerical simulation is necessary to understand the driving and evolution of the outflow during star formation and to determine star formation efficiency, as controlled by the protostellar outflow, in more detail.

Some authors have used a spherically symmetric calculation to investigate the evolution of a prestellar cloud until protostar formation  \citep{larson69,winkler80,masunaga98,masunaga00}.
Now, we can directly calculate the star formation from the prestellar cloud stage until protostar formation with a multi-dimensional calculation \citep{bate98,bate10,tomisaka02,banerjee06,machida07}. 
Before protostar formation, the first (adiabatic) core forms \citep{larson69} and drives a wide-opening-angle  outflow  \citep{tomisaka02,machida05b,hennebelle08b,tomida10,commerson10,burzel11}. 
After protostar formation, the first core evolves directly into the circumstellar disk after the protostar formation \citep{bate98,bate10,bate11,walch09a,machida10a,machida10c,machida11a}. 
%%In other words, the circumstellar disk originates in the first core.
Since the first core transitions smoothly to the circumstellar disk, the outflow driven by the first core before protostar formation is driven by the circumstellar disk after protostar formation without a transient disappearance.
In summary, a wide-opening-angle outflow appears before protostar formation and continues to be driven after protostar formation until the end of the main accretion phase \citep{machida08b,machida09a,machida10b}.
This type of outflow is believed to correspond to the molecular outflow frequently observed in star-forming regions \citep{bontemps96,wu04,arce10,curtis10}.
Recently, a bipolar molecular outflow was observed around the candidate first core Per-Bolo 58 \citep{dunham11}. 
In addition, some first core candidates were reported by several authors \citep{chen10,chen10b,enoch10}.
Thus, to estimate the mass ejection rate from the host cloud or star formation efficiency, we need to calculate the evolution of the protostellar outflow from the prestellar core stage because a wide-opening-outflow appears before  protostar formation.

In addition to the wide-opening-angle outflow, well collimated high-velocity jets are observed in the star forming region \citep{arce07, bally07}. 
Such a jet also appears during star formation process, and is driven in very close proximity to the protostar \citep{tomisaka02, banerjee06, machida08a, machida08b}.
The driving source and its spatial scale of each flow (wide-opening-angle outflow and well collimated jet) are considerably different: the driving source for the wide-opening-angle flow is the first core or the circumstellar disk, which  has a  size of $\gtrsim 1$\,AU, whereas the driving source for the well collimated jet is the protostar and the disk near the protostar in the region of $\ll 0.1$\,AU \citep{machida08b}.
In addition, the mass ejection rate of the wide-opening-angle flow is much higher than that of the well collimated  jet.
Because the jet has a good collimation, it cannot accumulate sufficient mass in the infalling envelope to reduce star formation efficiency \citep{tomisaka02,machida08b}.
Note that the difference in collimation between the wide-opening-angle outflow and well collimated jet is caused by the driving mechanism and the configuration of the magnetic field lines around the driving source \citep{machida08b}.

A statistical study of molecular outflows observed in various star forming regions also indicates that the (wide-opening-angle)  molecular outflows have a sufficient energy or momentum to control the star formation \citep{bontemps96,arce10}.
Therefore, the wide-opening-angle outflow and not the well collimated  jet is expected to be important in determining star formation efficiency.
Thus, to estimate the mass ejection rate (or star formation efficiency), we must resolve the driving source of the wide-opening-angle outflow, whereas we do not necessarily need to resolve the protostar itself, which drives the well collimated  jet.
Since previous studies spatially resolved the protostar, they could not directly estimate the resulting mass ejection rate by the protostellar outflow and star formation efficiency \citep{tomisaka02,banerjee06,machida08b}.
They calculated the evolution of the protostellar outflow only for a short duration of $\lesssim 1-10$\,yr at most, whereas the main accretion phase lasts for at least $\gtrsim 10^3-10^4$\,yr. 
It is  difficult to calculate the evolution of the protostellar outflow for a long duration by resolving the protostar itself because the timescales (time steps) in the regions around the protostar and the molecular cloud are quite different.

On the other hand, without resolving the driving source of the protostellar outflow, some studies focused on the long-term evolution of the protostellar outflow propagating into the interstellar medium.
In these studies, the outflow is input artificially in the computational domain \citep[e.g.,][]{arce07}.
In addition, the circumstellar disk, the configuration, and the strength of the magnetic field are also adopted arbitrarily.
Even with this type of calculation, we cannot estimate star formation efficiency because the mass ejection rate is artificially assumed in such studies.

At the expense of spatial resolution around the protostar, \citet{machida09a} calculated the evolution of the collapsing cloud from the prestellar stage until the end of the main accretion phase.
In their study, the driving source of the outflow (i.e., the first core and the circumstellar disk) was well resolved  spatially, whereas the protostar was not resolved and was replaced by sink cells.
They pointed out that the protostellar outflow reduces star formation efficiency to $\sim20-60\%$.
However, they investigated only the evolution of a low-mass cloud core  ($M\sim0.22\msun$) and did not investigate the evolution of a cloud with a typical mass scale of $\sim 1\msun$ \citep[e.g.,][]{motte98,onishi02}.
In this study, using conditions similar to those in \citet{machida09a}, we calculate the evolution of a cloud with various initial masses and investigate the evolution of the outflow and the impact of the protostellar outflow on star formation efficiency.
This paper is structured as follows. 
The framework of our models and the numerical method are described in \S 2. 
The numerical results are presented in \S 3. 
The mass ejection rate and star formation efficiency are discussed in \S 4.
We summarize our results in \S 5.

\section{Model and Numerical Method}
\label{sec:model}

\subsection{Basic Equations}
To investigate star formation efficiency and long-term evolution of the protostellar outflow, we calculate the star formation process from the prestellar core stage until the end of the main accretion phase for clouds with various masses using three-dimensional resistive MHD equations, including self-gravity:
\begin{eqnarray} 
& \dfrac{\partial \rho}{\partial t}  + \nabla \cdot (\rho \vect{v}) = 0, & \\
& \rho \dfrac{\partial \vect{v}}{\partial t} 
    + \rho(\vect{v} \cdot \nabla)\vect{v} =
    - \nabla P - \dfrac{1}{4 \pi} \vect{B} \times (\nabla \times \vect{B})
    - \rho \nabla \phi, & \\ 
& \dfrac{\partial \vect{B}}{\partial t} = 
   \nabla \times (\vect{v} \times \vect{B}) + \eta \nabla^2 \vect{B}, & 
\label{eq:reg}\\
& \nabla^2 \phi = 4 \pi G \rho, &
\end{eqnarray}
where $\rho$, $\vect{v}$, $P$, $\vect{B} $, $\eta$, and $\phi$ denote the density, velocity, pressure, magnetic flux density, resistivity, and gravitational potential, respectively.
To mimic the temperature evolution calculated by \citet{masunaga00}, we adopt the piece-wise polytropic equation of state \citep[see][]{vorobyov06,machida07} as\begin{equation} 
P =  c_{s,0}^2\, \rho \left[ 1+ \left(\dfrac{\rho}{\rho_c}\right)^{2/3} \right],
\label{eq:eos}
\end{equation}
 where $c_{s,0} = 190$\,m\,s$^{-1}$ and 
$ \rho_c = 3.84 \times 10^{-14} \, \rm{g} \, \cm$ ($n_c = 10^{10} \cm$). 
Equation~(\ref{eq:eos}) shows that the gas behaves isothermally for $n\lesssim10^{10}\cm$ and adiabatically for $n\gtrsim10^{10}\cm$.
For the realistic evolution of the magnetic field during star formation, we adopt a resistivity ($\eta$) of the fiducial value in \citet{machida07}, in which Ohmic dissipation becomes effective for $10^{11}\cm \lesssim n \lesssim 10^{15}\cm$ (for details, see Eqs.~[9] and [10] and Fig.~1 of \citealt{machida07}).

\subsection{Initial Settings}
\label{sec:setting}
This study investigates star formation efficiency in a single cloud.
Thus, we assume an isolated cloud core embedded in the interstellar medium.
As the initial state, we take a spherical cloud with a critical Bonnor--Ebert (BE) density profile $\rho_{\rm BE}$, in which a uniform density is adopted outside the sphere ($r > \rcri$, where $\rcri$ is the critical BE radius) to mimic the interstellar medium.
The gravitational force is ignored outside the BE sphere ($r>\rcri$) to avoid the inflow of gas  from outside the isolated core (i.e., from the interstellar medium).
In addition, to strictly avoid mass inflow from outside the core, we prohibit mass inflow at $r=R_c$.
Note that we do not prohibit mass outflow at the boundary between the BE sphere and the interstellar medium; the gas escapes freely from the BE sphere by the protostellar outflow.
Hereafter, we call the gravitationally bound gas cloud within $r<R_c$ the host cloud. 
We confirmed that the total mass of the host cloud is well conserved during the calculation before the protostellar outflow reaches the cloud boundary.
Note that the total mass inside the host cloud is not conserved after the protostellar outflow propagates into the interstellar space ($r>R_c$), because the mass is ejected from the host cloud by the protostellar outflow.

Since the critical BE sphere is in equilibrium, we increase the density by a factor of $f$ to promote contraction, where $f$ is the density enhancement factor that represents the stability of the initial cloud.
The density profile of the initial cloud is described as
\begin{eqnarray}
\rho(r) = \left\{
\begin{array}{ll}
\rho_{\rm BE}(r) \, (1+\delta_\rho)\,f & \mbox{for} \; \; r < R_{c}, \\
0.02\, \rho_{\rm BE}(R_c)\, (1+\delta_\rho)\, f  & \mbox{for}\; \;  r \ge R_{c}, \\
\end{array}
\right. 
\label{eq:dens}
\end{eqnarray}
where $\rho_{\rm BE}(r)$ is the density distribution of the critical 
BE sphere, and $\delta_\rho$ is the axisymmetric density perturbation. 
%%\begin{eqnarray}
%%\rho(r) = \left\{
%%\begin{array}{ll}
%%\rho_{\rm BE}(r) \, f & \mbox{for} \; \; r < R_{c}, \\
%%\rho_{\rm BE}(R_c)\, f & \mbox{for}\; \;  r \ge R_{c}. \\
%%\end{array}
%%\right. 
%%\end{eqnarray}
An initial cloud with larger $f$ is more unstable against gravity.
The cloud stability is generally represented by a parameter $\alpha_0$ ($\equiv E_{\rm t}/E_{\rm g}$), which is the ratio of thermal ($E_{\rm t}$) to gravitational ($E_{\rm g}$) energy.
As shown in \citet{matsu03}, when the BE density profile is adopted, the density enhancement factor is related to the parameter $\alpha_0$ as 
\begin{equation}
\alpha_0 = \dfrac{0.84}{f}.
\label{eq:alpha}
\end{equation}
We adopt a density enhancement factor as  $1.68$, which corresponds to $\alpha_0=0.5$. 
The density contrast between the center of the cloud ($r=0$) and the cloud boundary ($r=R_{\rm c}$) is $\rho (r=0)/\rho (r=R_{\rm c}) = 14$.
In addition, a uniform density of $\rho_{\rm amb}=0.02\rhoc$ (i.e, $2\%$ of the central cloud density) is adopted outside the sphere ($r > \rcri$).

To break the axial symmetry, we add a small amount of $m=2$-mode non-axisymmetric density perturbation to the initial core.
For the $m=2$-mode, in equation~(\ref{eq:dens}), we choose
\begin{equation}
\delta_\rho = A_{\phi} (r/R_{\rm c})^2\, {\rm cos}\, 2\phi, 
\label{eq:dens-pert}
\end{equation}
where $A_{\phi}$ (=0.01) represents the amplitude of the perturbation.
The radial dependence is chosen so that the density perturbation remains regular at the origin ($r = 0$) at one time-step after the initial stage.
This perturbation develops into a non-axisymmetric perturbation in the circumstellar disk that contributes to  angular momentum transfer.
In addition, this $m=2$ perturbation ensures that the center of gravity is always located at the origin.

For a dimensional BE density profile, we adopt an isothermal temperature of $T=10$\,K and a central number density of $\nc =  n_{c,0}$, where $n_{c,0}$ is a parameter in the range  $n_0 = 3\times10^5 - 3 \times 10^9\,f\,\cm$.
Since the temperature of each initial cloud is fixed, the size (i.e., radius and mass) of the BE sphere is uniquely determined only by the parameter $n_{c,0}$.
Thus, each model is characterized only by the initial central density, $n_{\rm c,0}$.
With these parameters, the critical BE radius (or radius of the host cloud) is  $R_{\rm c} = 87-8700$\,AU.
The mass inside $r < \rcri$ for each model is  $M_{\rm cl}=0.015-1.5\msun$.
The host cloud radius and mass for each model are listed in Table~\ref{table:1}.

In each model, the cloud rotates rigidly around the $z$-axis in the  $r< \rcri$ region and a uniform magnetic field parallel to the $z$-axis (or rotation axis) is adopted in the entire computational domain.
The magnetic field strength and the rotation rate are scaled using the central density $\rho_0 = \rho_{\rm BE}(r=0) f$ as
\begin{equation}
\alpha =  B_0^2 / (4\pi \, \rho_0 \, c_{s,0}^2),
\label{eq:alpha}
\end{equation}
\begin{equation}
\omega = \Omega_0/(4 \pi\,  G \, \rho_0  )^{1/2}.
\label{eq:omega}
\end{equation}
In all models, we adopt $\alpha=0.1$ and $\omega=0.1$, which are the most appropriate parameters for driving strong outflow in the collapsing cloud \citep{machida05b,machida08b}.
Since the magnetic field and rotation are normalized by the central density and each model is characterized by only the central density, each cloud has a different magnetic field strengths and angular velocity.
The magnetic field ($B_0$) and angular velocity ($\Omega_0$) for each model are summarized in Table~\ref{table:1}.
However, all models have the same ratios of rotational and magnetic energies to the gravitational energy, $\beta_0$ ($\equiv E_{\rm rot}/E_{\rm grav} =5\times10^{-3}$) and $\gamma_0$ ($\equiv E_{\rm mag}/E_{\rm grav}=4\times10^{-2}$), where $E_{\rm rot}$ and $E_{\rm mag}$ are rotational and magnetic energies, respectively.
In addition, all models have the same mass-to-flux ratio $M/\Phi$.
There exists a critical value of $M/\Phi$ below which a cloud is supported against  gravity by the magnetic field.
For a cloud with uniform density,  \citet{mouschovias76} derived a critical mass-to-flux ratio
\begin{equation}
\left(\dfrac{M}{\Phi}\right)_{\rm cri} = \dfrac{\zeta}{3\pi}\left(\dfrac{5}{G}\right)^{1/2},
\label{eq:mag2}
\end{equation}
where the constant $\zeta=0.48$ \citep{tomisaka88a,tomisaka88b}.
The mass-to-flux ratio normalized by the critical value $\lambda$ is described as
\begin{equation}
\lambda  \equiv \left(\dfrac{M}{\Phi}\right) \left(\dfrac{M}{\Phi}\right)_{\rm cri}^{-1}.
\label{eq:crit}
\end{equation}
All models have $\lambda=4$, which is  slightly larger than the typical value of observation.
The observations indicate that molecular cloud cores have the mass-to-flux ratio in the range of $0.8\lesssim \lambda \lesssim7.2$ with a median value of $\lambda \approx 2$ \citep{crutcher99}.

The model names and parameters are also listed in Table~\ref{table:1}.
We believe that these energy ratios are adequate for comparing cloud evolution among models with different masses.
We can match the magnetic field strength ($B_0$) and rotation rate ($\Omega_0$) for clouds with different masses. 
However, since doing so changes the evolution of the cloud and the protostellar outflow, it is difficult to compare the star formation efficiency of the  models.
For example, when the initial magnetic field strength is fixed among the models, no outflow may appear in a less massive cloud because it has a small ratio of magnetic energy to gravitational energy.
%%fragmentation may occurs in cloud with large mass that have larger rotational energy toward gravitational energy, while no fragmentation may occur in small mass cloud that have smaller rotation energy. 
%%The cloud evolution for cloud with different rotational and magnetic energies is investing in a subsequent paper.
%%we think that not the absolute value of the magnetic field and rotation rate but the energy ratio (i.e., $\alpha_0$, $\beta_0$ and $\gamma_0$) should be matched.

In this paper, to compare cloud evolution among models, we usually use the freefall timescale at the center, $r=0$ ($\tffc$), and boundary, $r=R_c$ ($\tffb$), of the initial cloud as the unit of time.
Since the density contrast between the center of the cloud and its boundary is 14, the freefall timescale of the cloud boundary is about 3.7 times longer than that at the center of the cloud ($\tffb=3.7\tffc$).
The freefall timescale at the center of the initial cloud ($\tffc$) for each model is listed in Table~\ref{table:1}.
%%\clearpage
%%\subsection{Numerical Method}

%%Each model has different grid sizes and cell widths.
%%For models 1-6, box size of the first level of grid ($l=1$)  is  $L_1 \sim 2.1\times 10^5$\,AU, and the maximum grid level is $l_{\rm max} = 9$ that has a box size of $L_9=820$\,AU and a cell width of $\Delta_{l=9}= 13$\,AU.
%%On the other hand,  model 7 has $L_1  \sim 2.1\times 10^6$\,AU and $l_{\rm max} = 13$ that has a box size of $\Delta_{l=13}=507$\,AU and a cell width of $\Delta_{l=13}=8$\,AU.

\subsection{Sink Cell and Numerical Method}
\label{sec:sink}
To realize the long-term calculation of star formation, we adopt a sink at the center of the cloud.
We start the calculation without a sink and calculate the cloud evolution for the prestellar gas collapse phase without a sink.
Later, we identify protostar formation in the collapsing cloud core when the number density exceeds $n > n_{\rm thr}$ at the cloud center, where $n_{\rm thr}$ is the threshold density.
After protostar formation, we calculate the cloud evolution with the sink.

To model the protostar, we adopt a fixed sink with a radius of $1\,$AU composed of sink cells only around the center of the computational domain.
Since we add only $m=2$ density perturbation, as described in \S\ref{sec:setting}, the protostar (or center of gravity) does not move and remains at the center of the computational domain during the calculation.
In the region $r < r_{\rm sink} = 1\,$AU, the gas having a number density of $n > n_{\rm thr} = 10^{13}\cm$ is removed from the computational domain and added to the protostar as a gravitating mass in each time step \citep[for details, see][]{machida09a}.
Thus, for each time step, the accretion mass onto the protostar is calculated as
\begin{equation}
M_{\rm acc} = \int_{r < r_{\rm sink}} [\rho(i,j,k) - \rho_{\rm thr}]\, dV.
\end{equation}
In addition, inside the sink, the magnetic flux is removed by Ohmic dissipation because this region has a magnetic Reynolds number $Re$ exceeding unity $Re>1$ \citep[for details, see][]{machida07}.

For calculation on a large spatial scale, the nested grid method is adopted \citep[for details, see ][]{machida05a,machida05b}. 
Each level of a rectangular grid has the same number of cells ($ 64 \times 64 \times 32 $).
The calculation is first performed with five grid levels ($l=1-5$).
In all models, the fifth level of the grid ($l=5$) has a box size of $L_5=2\,\rcri$ and just covers the entire  region of the isolated BE sphere.
The  first level of the grid has a box size of $L_1=2^5\,\rcri$ and is filled with low-density interstellar medium outside $r>\rcri$.
Thus, we can calculate the propagation of the protostellar outflow in the region of  $<2^5\,\rcri$.
The protostellar outflow never reaches the computational boundary by the end of the calculation.
In addition, we have checked that the Alfv$\acute{\rm e}$n waves generated at the center of the cloud (or the computational boundary) never reaches the computational boundary (or the center of the cloud) during the calculation \citep[for details, see][]{machida10b}.
After the calculation starts, a new finer grid is generated before the Jeans condition is violated \citep{truelove97}.
Although the maximum grid level differs among the models, each model has a spatial resolution of $<0.3$\,AU in the finest grid.

\section{Results}
In this section, we present the evolution of the collapsing cloud and the outflow for typical models in \S\ref{sec:typical} (model N08, less massive cloud case) and \S\ref{sec:typical2} (model N06, massive cloud case). 
Then, we describe the mass accretion rate (\S\ref{sec:accretion}) and masses of the protostar and circumstellar disk (\S\ref{sec:mass}), and the mass ejected by the protostellar outflow (\S\ref{sec:out}).
The properties of the protostellar outflow are described in \S\ref{sec:out2}.

\subsection{Typical Model}
\subsubsection{Model N08}
\label{sec:typical}
\begin{figure}
\includegraphics[width=150mm]{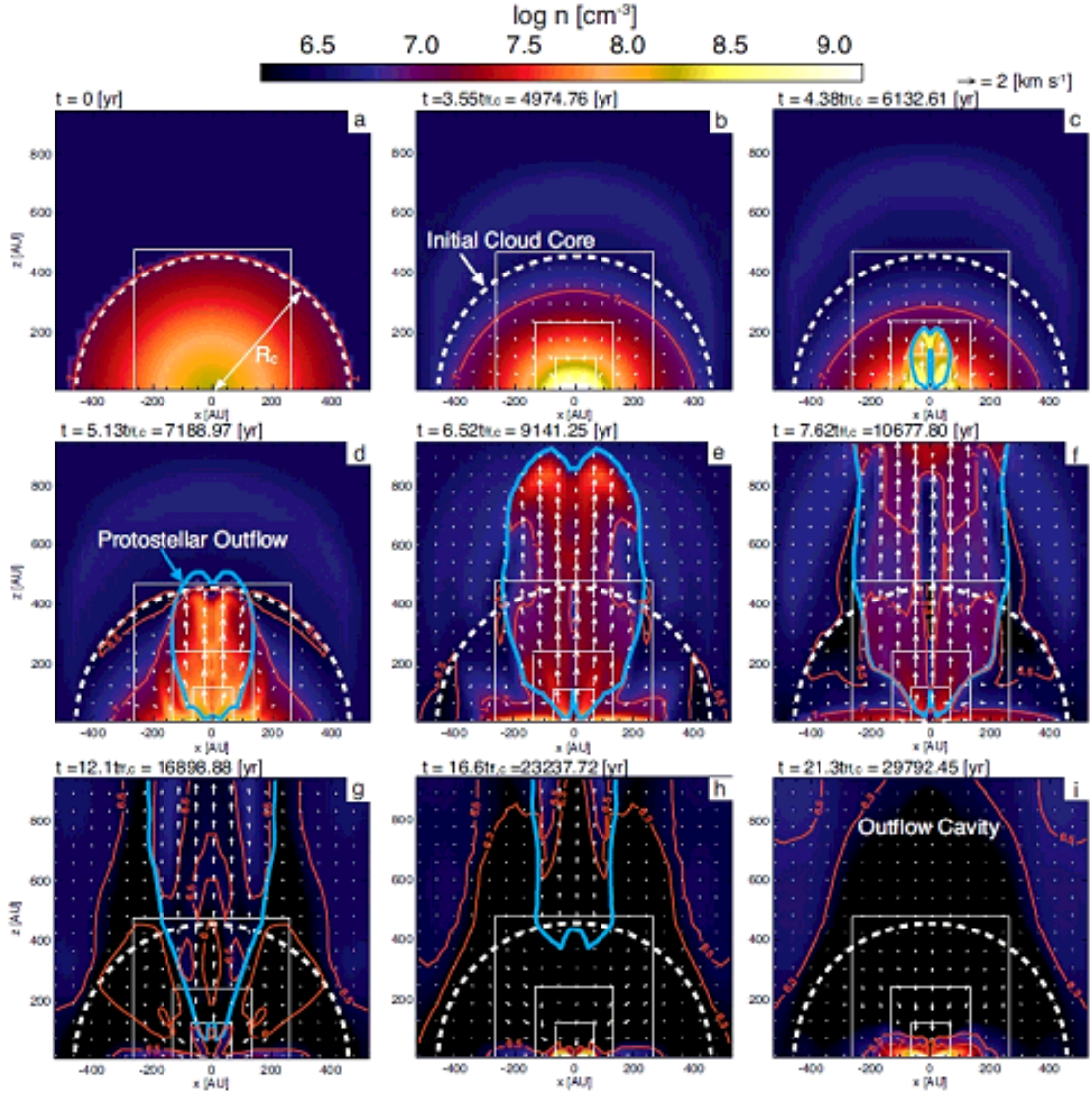}
\caption{
Time sequence images  from the initial state until the end of the main accretion phase for model N08.
In each panel, the density (color and red contours) and velocity (arrows) distribution on the $y=0$ plane are plotted with the initial cloud scale.
The white dashed circle represents initial cloud radius (i.e., the host cloud).
The blue line is the boundary of the outflow inside which the gas moves outwardly toward the center of the cloud (or the protostar) with a supersonic velocity.
The elapsed time $t$ in unit of the freefall timescale ($\tffc$) and year is plotted on the upper side of each panel.
The white squares in each panel denote the outer boundary of the subgrid.
}
\label{fig:1}
\end{figure}

Figure~\ref{fig:1} plots the cloud evolution for model N08 from the initial state until the end of the main accretion phase; only the $z>0$ region is shown.
Note that although the structure only around the initial cloud scale ($\sim R_{\rm c}$) is plotted in Figure~\ref{fig:1}, the computational domain has a size of $2^5 R_{\rm c}$.
As seen in Figure~\ref{fig:1}{\it a}, we adopted a spherical cloud with the BE density profile and the radius of $R_{\rm c}$ as the initial state.
As described in \S\ref{sec:setting}, since we ignored the gravity outside the BE sphere ($r>R_{\rm c}$), only the gas in the $r<R_{\rm c}$ region (inside the white dotted line in Fig.~\ref{fig:1}) can collapse to fall onto the center of the cloud.
For this model, the number density exceeds $n > n_{\rm thr}$ and a protostar forms at $t=3.51\tffc$, where $\tffc$ ($=1.4\times10^3$\,yr) is the freefall timescale of the initial cloud at the center.
Figure~\ref{fig:1}{\it b} shows the density distribution just after protostar formation.
The red contours of $n=10^7\cm$ in  Figure~\ref{fig:1}{\it a} and {\it b} indicate that the cloud gradually contracts toward the center with time.

The blue contour in the figure corresponds to the boundary of the protostellar outflow, inside which the gas moves outward against the center of the cloud.
Thus, this contour shows the shape of the protostellar outflow.
The outflow is driven by the circumstellar disk that originates from the first core, which appears before protostar formation \citep{tomisaka02,machida05b,hennebelle08a,burzel11}.
After protostar formation, the first core becomes the circumstellar disk \citep{bate98,inutsuka10,machida10a}.
The outflow is then driven by the circumstellar disk after a smooth transition from the first core 	to the circumstellar disk \citep{machida09a,machida11a}.
The protostellar outflow remains in the host cloud ($r<R_{\rm c}$) for $t\lesssim 7000$\,yr (Fig.~\ref{fig:1}{\it c}); it reaches the boundary of the host cloud at $t\sim7200$\,yr (Fig.~\ref{fig:1}{\it d}) and penetrates the host cloud boundary  and propagates into the interstellar space ($r>R_{\rm c}$) for $t\gtrsim 7200$\,yr (Fig.~\ref{fig:1}{\it e}-{\it h}).
The outflow driving halts inside the host cloud at $t\simeq17000$\,yr ($\simeq 12 \tffc$).
The infalling envelope is depleted and the mass accretion is almost over by this epoch (\S\ref{sec:mass}).
Thus, the circumstellar disk cannot drive the outflow at this epoch because the outflow is powered by gas accretion onto the circumstellar disk.
Since the freefall timescale of the cloud boundary is $\tffb=5300$\,yr, a large fraction of the gas inside the host cloud has already fallen onto either the circumstellar disk or the protostar by this epoch.
Although the outflow at the host cloud scale completely disappears in $t\sim3\times 10^4$\,yr (=$20$\,$\tffc$ = $5.4\,\tffb$),  the density cavity formed by the protostellar outflow remains around the host cloud as seen in Figure~\ref{fig:1}{\it i}.

Figure~\ref{fig:1} also shows that the infalling envelope in the host cloud depletes with time.
Part of the gas inside the initial host cloud is ejected by the protostellar outflow, and the remainder falls onto either the circumstellar disk or the protostar.
In Figures~\ref{fig:1}{\it h} and {\it i}, we can see only a disk-like structure with the size of $\sim$200\,AU, because the infalling gas is depleted by this epoch.
At these epochs, the density of the infalling envelope ($n \lesssim 10^5 \cm$)  is  less than $1/100$ the initial cloud density ($n\gtrsim 10^7\cm$).
At the end of the calculation ($t\simeq 20\,\tffc$), the mass ratio of the infalling envelope to the total mass of the initial cloud is only $<4\%$.
Thus, the main accretion phase has already ended by this epoch.

\begin{figure}
\includegraphics[width=150mm]{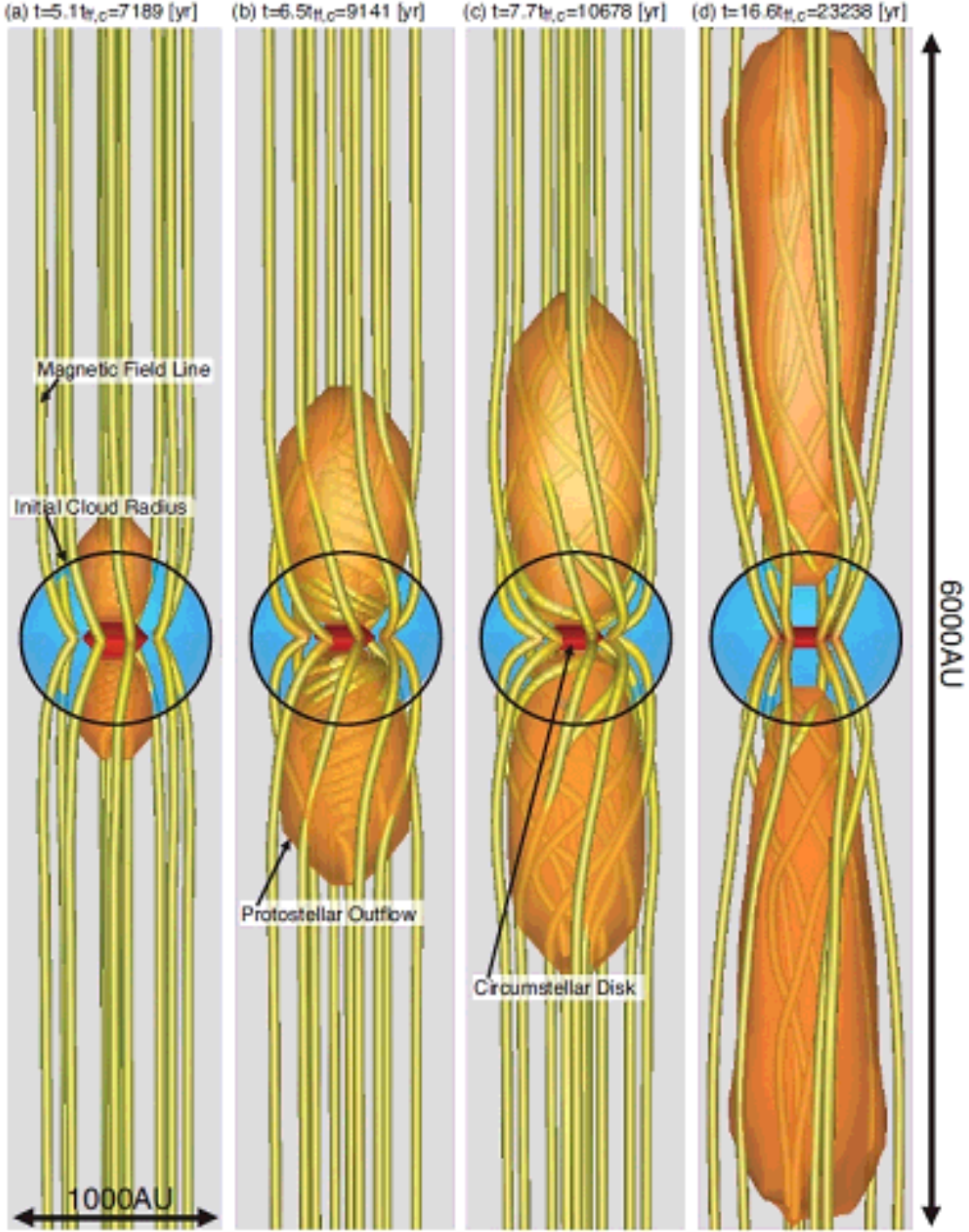}
\caption{
The magnetic field lines (yellow lines), outflow shape (orange iso-velocity surface) and high-density gas region  (central red iso-volume) are plotted in three-dimensions.
The orange surface is the isovelocity surface of $v_r = c_s$ inside which the gas is outflowing from the center of the cloud with the supersonic velocity $v_r > c_s$.
The blue sphere corresponds to the initial host cloud.
The elapsed time $t$ in unit of the freefall timescale ($\tffc$) and year is plotted on the upper side of each panel.
}
\label{fig:2}
\end{figure}

The evolution of the outflow configuration is shown in Figure~\ref{fig:2}.
Each panel corresponds to the same epoch as in Figure~\ref{fig:1}{\it d}, {\it e}, {\it f}, and {\it h}.
Figure~\ref{fig:2} shows that the protostellar outflow propagates into the interstellar space while maintaining good collimation.
However, Figure~\ref{fig:2}{\it a}  and {\it b} indicate that, early in the evolution, the outflow also extends in a transverse direction and increases in width.
After its width becomes comparable to the host cloud scale, the outflow extends only in the vertical direction while maintaining its width.
The magnetic field lines that guide the outflow are anchored by the host cloud (or the gravitationally bound sphere).
Thus, the final width of the outflow is comparable to that of the host cloud.
%%it is natural that outflow finally has a width comparable to  host cloud.
Therefore, the outflow collimation improves with time after the outflow width becomes comparable to the host cloud scale, whereas the collimation is not good when its width is smaller than the host cloud scale or the outflow remains within the host cloud.

\subsubsection{Model N06}
\label{sec:typical2}
\begin{figure}
\includegraphics[width=150mm]{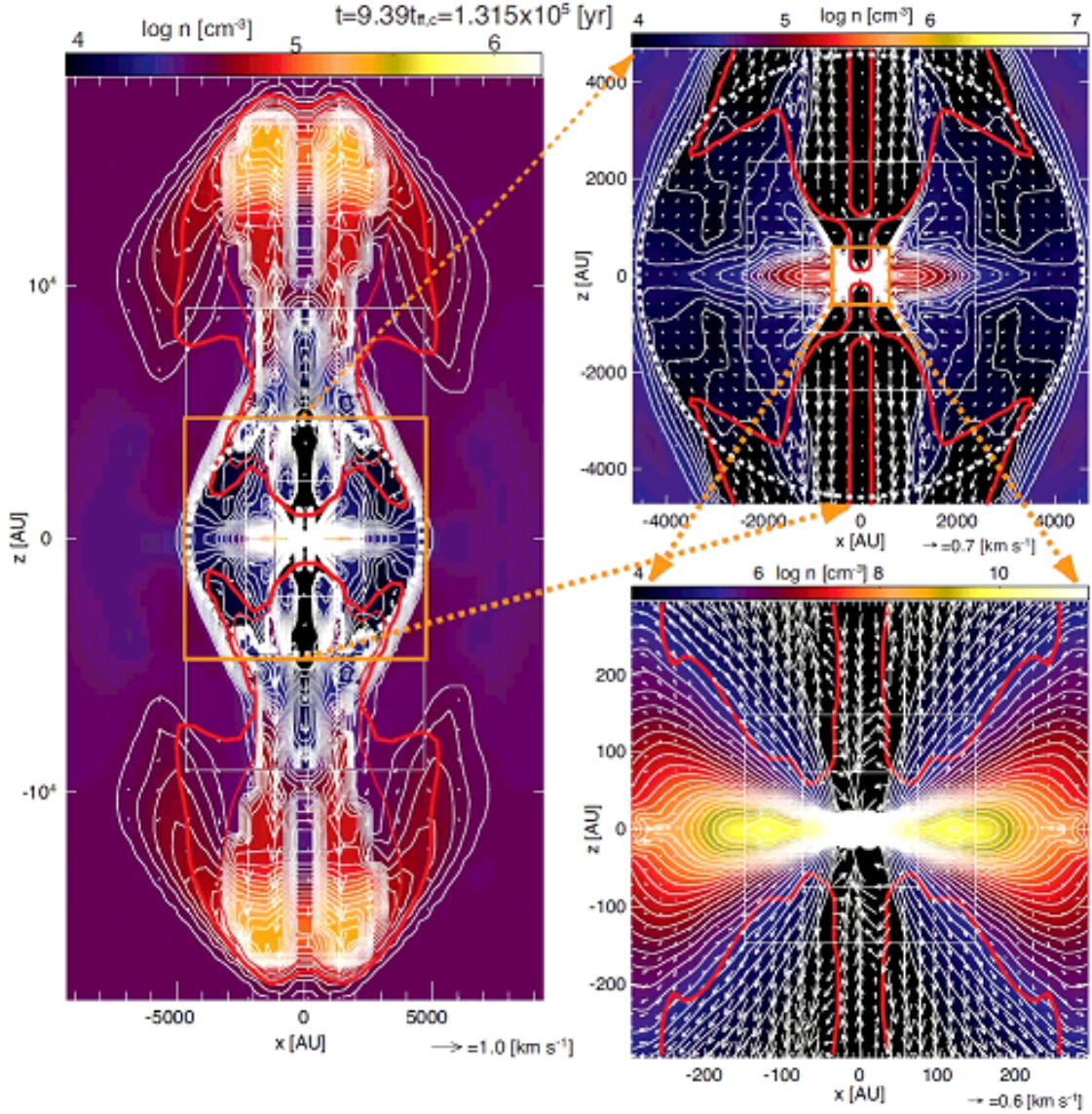}
\caption{
Density (color) and velocity (arrows) distributions on the $y=0$ plane at $t=9.39\,\tffc$ for model N06 with different spatial scales.
The white dashed circle represents initial cloud radius (i.e., BE radius).
The red contours denote the outflow inside which the gas moves outwardly toward the center of the cloud with a supersonic velocity (thick line) and half of the supersonic velocity (thin line).
}
\label{fig:3}
\end{figure}

\begin{figure}
\includegraphics[width=150mm]{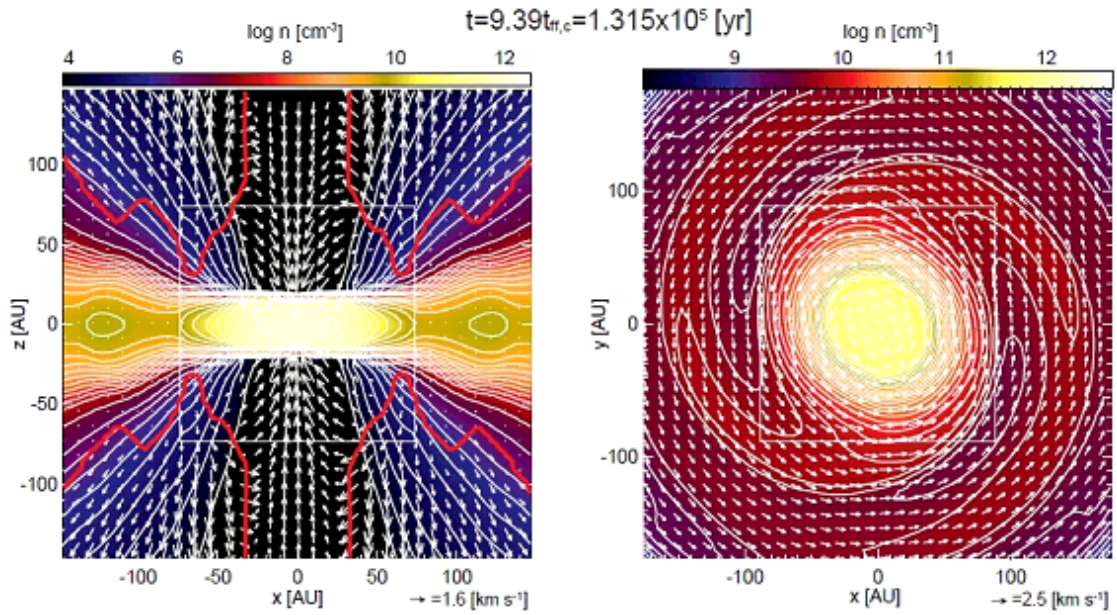}
\caption{
Density (color) and velocity (arrows) distributions on the $y=0$ (left) and $z=0$ (right) planes at the same epoch as in Fig.~\ref{fig:3}.
The red contour denotes  the outflow inside which the gas moves outwardly toward the center of the cloud with a supersonic velocity.
}
\label{fig:4}
\end{figure}

Figures~\ref{fig:3} and \ref{fig:4} show the density and velocity distributions for model N06 at $t=1.315\times10^5$\,yr ($=9.4\,\tffc$=$2.5\,\tffb$).
For this model, the protostar forms at $t=4.502\times10^4$\,yr after the cloud begins to collapse.
Thus, the figures show the structure  $8.647\times10^4$\,yr after the protostar formation.
By this epoch, the outflow penetrated the host cloud that has a radius of 4800\,AU and reaches $\sim2\times10^4$\,AU from the center of the cloud, as seen in Figure~\ref{fig:3} left panel.
The figure also shows the bow shocks caused by the protostellar outflow  at $z\simeq \pm 2\times10^4$\,AU.
The upper right panel in Figure~\ref{fig:3} plots the structure around the host cloud and shows that the gas  flows out from the host cloud in the vertical direction by the protostellar outflow.
In addition, inside the white dotted line that denotes the size of the initial cloud, the gas density is considerably lower than the ambient medium.
This is because a part of the gas falls onto the central region to form the protostar and circumstellar disk, while the remainder is ejected from the host cloud by the outflow.
The lower right panel is 16 times magnified view of the central region of the upper right panel.
This panel shows that the outflow is strongly driven by the disk-like structure around the protostar.
On this scale, the gas accretes onto the protostar through the circumstellar disk, and a part of the accreting mater is ejected by the outflow.
Figure~\ref{fig:4} shows the structure around the circumstellar disk on $y=0$ (left panel) and $z=0$ (right panel) planes.
These panels show that the rotating disk exits around the protostar and drives the outflow.
Thus, it is expected that the outflow is mainly driven by the magnetocentrifugal mechanism \citep{blandford82}.

Figures~\ref{fig:3} and \ref{fig:4} indicate that large scale outflow with a size of $>10^4$\,AU (Fig.~\ref{fig:3} left panel) originates from the disk wind driven by the circumstellar or rapidly rotating disk with a size of $\sim1-100$\,AU.
The disk wind propagates into the infalling envelope, and thus it sweeps and collects a larger fraction of the infalling matter.
Finally, a large fraction of the swept material is expelled from the host cloud.
Thus, the outflow significantly affects the final protostellar mass or star formation efficiency.

\subsection{Mass Accretion Rate onto Protostar}
\label{sec:accretion}
\begin{figure}
\includegraphics[width=150mm]{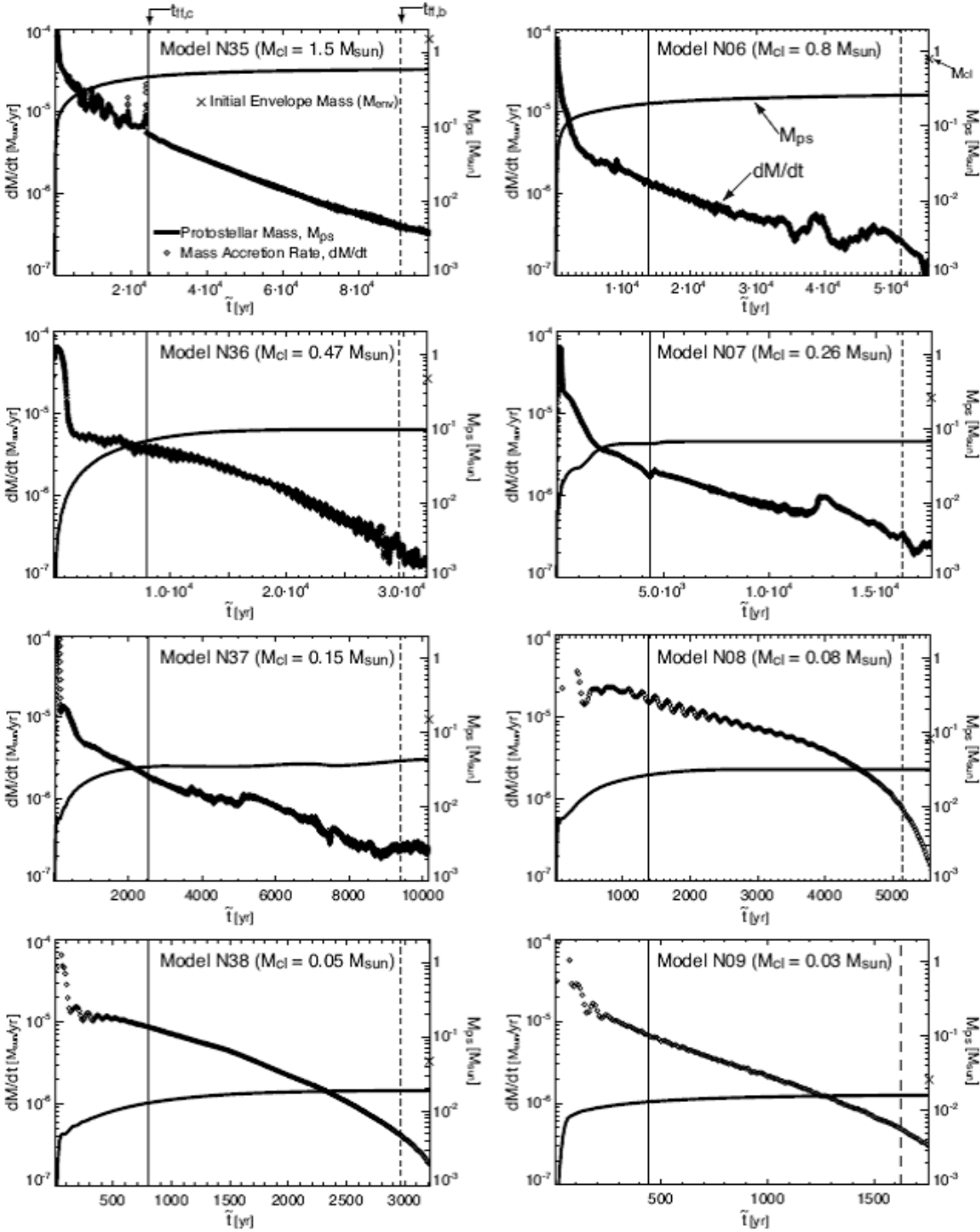}
\caption{
The mass accretion rate (left axis; diamond symbol) and protostellar mass (right axis; thick line) are plotted against the time $\tilde{t}$ (=$t-t_0$, where $t_0$ is time at the protostar formation) after the protostar formation for each model.
The vertical lines in each panel corresponds to  the freefall timescale of the host cloud at the center ($\tffc$, solid line) and cloud boundary ($\tffb$, broken line).
The initial host cloud mass is plotted by the cross ($\times$) on the right axis.
}
\label{fig:5}
\end{figure}

Figure~\ref{fig:5} shows the mass accretion rate onto the protostar and the protostellar mass against the time after protostar formation $\tilde{t}$.
Here, we describe the time after the protostar formation as $\tilde{t}$, which is defined as
\begin{equation}
\tilde{t} \equiv t-t_0,
\end{equation}
where $t$ is the elapsed time after the cloud begins to collapse (or the calculation starts), and  $t_0$ is the protostar formation epoch and  is listed in Table~\ref{table:1}.
In each model, the gas density exceeds $n>n_{\rm thr}$, and the protostar forms $\sim3-5\,\tffc$ after the cloud begins to collapse (Table~\ref{table:1}). 
Note that model N39 shows no continuous collapse and  no protostar appears; this is because the initial cloud density for model N39 is too high to induce continuous collapse.
In our models, since the gas becomes adiabatic at $n=n_c \sim 10^{10}\cm$ (\S\ref{sec:setting}),  the initial cloud density ($\nc=3\times 10^9\,f\,\cm$) for model N39 is very close to this critical value $n_c$. 
As a result, model N39 shows repeated contraction and expansion around the initial configuration, not continuous collapse.
Below, we describe only the models showing protostar formation (N35, N06, N36, N07, N37, N08, N38, N09) and do not mention model N39 again.

As described in \S\ref{sec:sink}, we removed the gas having a number density of $n>n_{\rm thr}$ in the region $r<1$\,AU from the computational domain.
We regarded the removed gas as the accreted mass onto the protostar and estimated the mass accretion rate in each time step.
Figure~\ref{fig:5} shows that, in each model, the mass accretion rate is as high as $\dot{M} \sim10^{-4}-10^{-5}\mdot$ just after protostar formation ($\tilde{t} \simeq0$\,yr). 
In theoretical analyses, the mass accretion rate is defined as $\dot{M}=f\, c_s^3/G $, where $f$ is a numerical factor (e.g., $f=0.975$ in \citealt{shu77}, $f= 46.9$ in \citealt{hunter77}).
Since gas clouds have temperatures of $T=10$\,K ($c_s=0.2\km$), the accretion rate in the main accretion phase is $\dot{M}=(2-90)\times 10^{-6}\mdot$.
Thus, the accretion rate derived in our calculations corresponds well to the theoretical expectation.

Then, in each model, the mass accretion rate gradually decreases with time, to $\dot{M} \sim10^{-5}-10^{-6}\mdot$ at $\tilde{t} \simeq \tffc$ and $\dot{M} \sim10^{-6}-10^{-7}\mdot$ at $\tilde{t} \sim \tffb$.
Thus, gas accretion almost halts and the protostar rarely increases in mass at $\tilde{t} \gtrsim \tffb$.
As shown in Figure~\ref{fig:5}, all models show a qualitatively and quantitatively similar mass accretion rate trend when the mass accretion rate onto the protostar is normalized by the freefall timescale of the initial cloud. 
However, since the real (or dimensional) time of the freefall timescale depends on the initial cloud density (or the initial cloud mass), the duration of the main accretion phase is different in models with different cloud masses.  
A cloud with a larger mass (or lower density) has a longer duration of the main accretion phase.
For example, model N35 with $M_{\rm cl}=1.5\msun$ has the mass accretion of $\dot{M}>10^{-6}\mdot$ for $\sim5\times10^4$\,yr, while model N09 with $M_{\rm cl}=0.03\msun$ has that only for $\sim 10^3$\,yr.
Figure~\ref{fig:5} also shows that the protostar formed in a massive cloud is more massive than that formed in a less massive cloud.
This is because the massive cloud has a longer gas accretion phase, so the protostar has enough time to acquire sufficient mass and evolves into a relatively massive star.

In Figure~\ref{fig:5}, the mass accretion rate oscillates in models having an initially massive cloud.
This is caused by a non-axisymmetric structure appearing in the circumstellar disk.
Since the circumstellar disk is more massive than the protostar just after protostar formation \citep{bate98, walch09a, machida10a,tsukamoto11}, a non-axisymmetric or spiral structure develops owing to gravitational instability.
Such structure effectively transfers angular momentum outward and intermittently promotes mass accretion onto the protostar. 
As a result, these models show time variability in the mass accretion rate \citep{vorobyov06, machida11a}. 
In Figure~\ref{fig:5}, models N35, N06, N36, N07 and N37 show time variability in the mass accretion rate.

\subsection{Mass Evolution of Protostar and Circumstellar Disk}
\label{sec:mass}
\begin{figure}
\includegraphics[width=150mm]{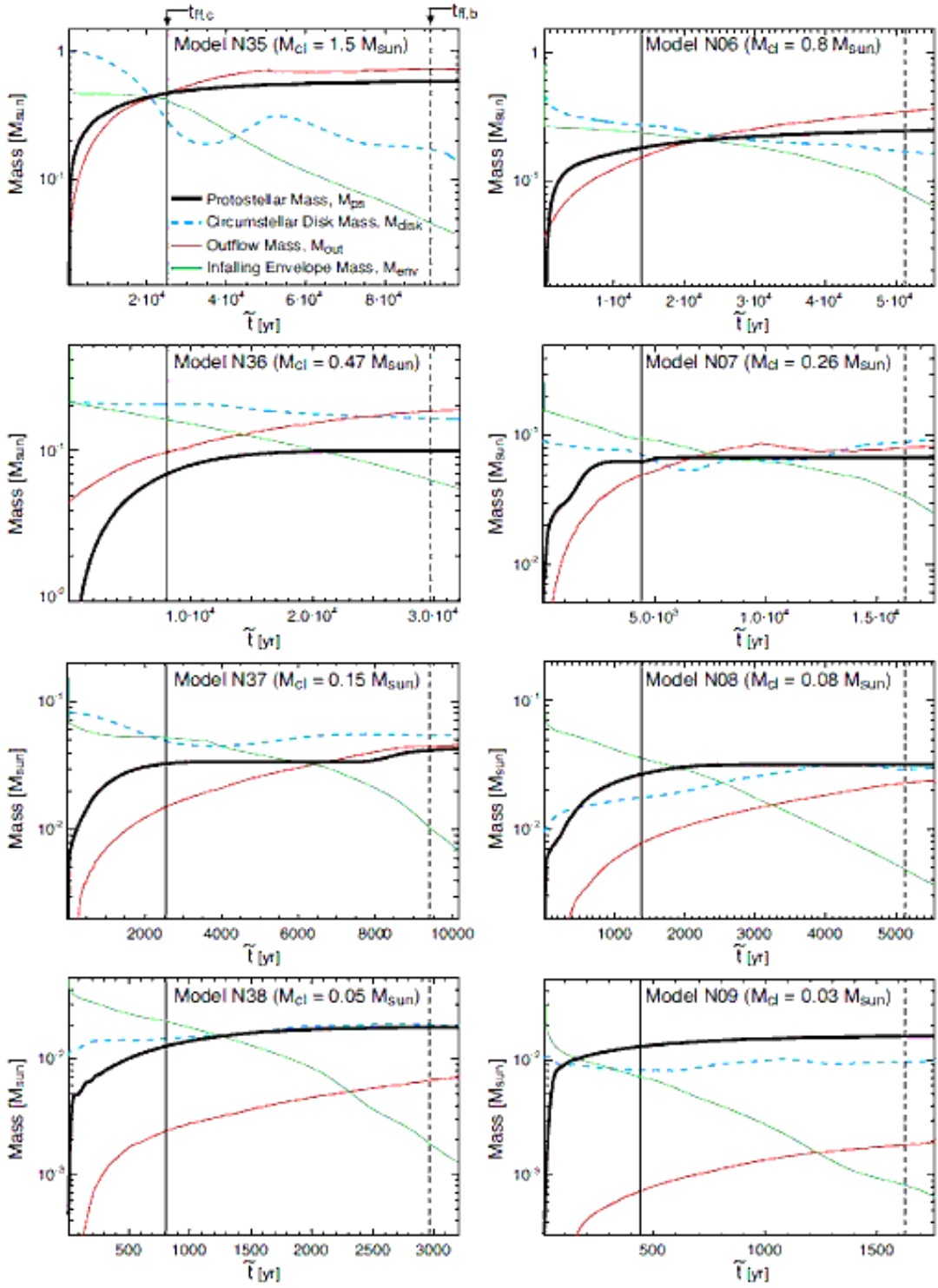}
\caption{
The mass of protostar, circumstellar disk, outflow and infalling envelope is plotted against the time $\tilde{t}$ (= $t-t_0$) after the protostar formation for each model.
The vertical lines in each panel corresponds to  the freefall timescale of the host cloud at the center ($\tffc$, solid line) and cloud boundary ($\tffb$, broken line).
}
\label{fig:6}
\end{figure}

The mass of the protostar, circumstellar disk, protostellar outflow, and infalling envelope are plotted against time after protostar formation $\tilde{t}$ in Figure~\ref{fig:6}.
In addition, these masses at $\tffb$ after protostar formation are listed in Table~\ref{table:2}.
The circumstellar disk mass is estimated according to the formula in \citet{machida10a}.
To estimate the outflowing gas $M_{\rm out}$, we integrated the gas with velocity $v_r > c_{\rm s}$ for the entire computational domain and subtracted the gas swept by the protostellar outflow outside the host cloud $r>R_c$ from the integrated mass.
For later convenience, we divide the outflowing mass into two parts, 
\begin{equation}
M_{\rm out} = M_{\rm ej} + M_{\rm out, R_c},
\end{equation}
where $M_{\rm ej}$ is the mass ejected from the host cloud, and $M_{\rm out, R_c}$ is the outflowing mass having $v_r>c_{\rm s}$ inside the host cloud ($r<R_{\rm c}$).
To calculate the mass of the infalling envelope, we calculated the total mass $M_{\rm tot}$ inside the host cloud ($r<R_c$).
Then, we subtracted the disk mass $M_{\rm disk}$ and the outflowing mass $M_{\rm out, R_c}$ in the  $r<R_c$ region from the total mass,
\begin{equation}
M_{\rm env} = M_{\rm tot} (r<R_c) - M_{\rm disk} - M_{\rm out, R_c} (r<R_c).
\end{equation}
In our models, the sum of the protostellar mass, circumstellar mass, and infalling envelope mass is not conserved inside the host cloud (i.e., $r<R_c$) because part of the gas is ejected from the host cloud by the protostellar outflow.
However, we confirmed that the sum of the total mass and the protostellar mass is well conserved before the outflow reaches the boundary of the host cloud,  as described in \S\ref{sec:setting}.
%%In addition, in our previous work \citep{machida10a}, we have checked that the mass conservation is satisfied  with 5\% accuracy without magnetic field (i.e., without protostellar outflow).
%%Thus, we can safely estimate each mass in our calculation.

Figure~\ref{fig:6} shows that, in each model, the mass of the infalling envelope decreases to $M_{\rm env}/M_{\rm cl} \lesssim 0.1$ at $\tilde{t} \sim \tffb$ (Table~\ref{table:2}), where $M_{\rm cl}$ is the initial cloud mass.
Thus,  the mass of the infalling envelope is depleted and the main accretion phase is almost over by this epoch.
Note that, in Figure~\ref{fig:6}, we subtracted the time ($t_0$) until the protostar forms from the time ($t$) after the cloud begins to collapse.
Thus, at the epoch $\tffb$, indicated by the arrow in Figure~\ref{fig:6}, a longer time than $\tffb$  has passed since the cloud begins to collapse (Table~\ref{table:1}).
Therefore, it is reasonable that almost all the gas has already fallen onto the center of the cloud at this epoch $\tilde{t}\sim \tffb$, because the gas falls onto the cloud center in several times freefall timescale \citep{machida05a} and the freefall timescale at the cloud boundary ($\tffb$) has already passed by this epoch.

Figure~\ref{fig:6} also indicates that the circumstellar disk mass dominates the protostellar mass just after protostar formation in each model.
The circumstellar has been reported to originate from the first core \citep{inutsuka10}, which is about 10-100 times more massive than the protostar at the protostar formation epoch \citep{larson69,masunaga00}.
\citet{machida10a} showed that the circumstellar disk is inevitably more massive than the protostar in the early main accretion phase because the first core evolves directly into the circumstellar disk after protostar formation.
In the later main accretion phase, however, in some models, the protostellar mass dominates the circumstellar disk mass.
In the initially massive clouds (models N35, N06, and N36), the protostellar mass continues to increase, whereas the circumstellar disk mass gradually decreases until the end of the main accretion phase.
In addition, in models N35 and N06, the protostar becomes more massive than the circumstellar disk for $t\gtrsim \tffc$.
The massive circumstellar disk becomes gravitationally unstable and tends to exhibit a non-axisymmetric structure that can transfer angular momentum outward; the gas in the circumstellar disk effectively falls onto the protostar.
Therefore, the accretion rate onto the protostar from the circumstellar disk dominates the accretion rate onto the circumstellar disk from the infalling envelope, and the mass of the disk begins to decrease.

\begin{figure}
\includegraphics[width=150mm]{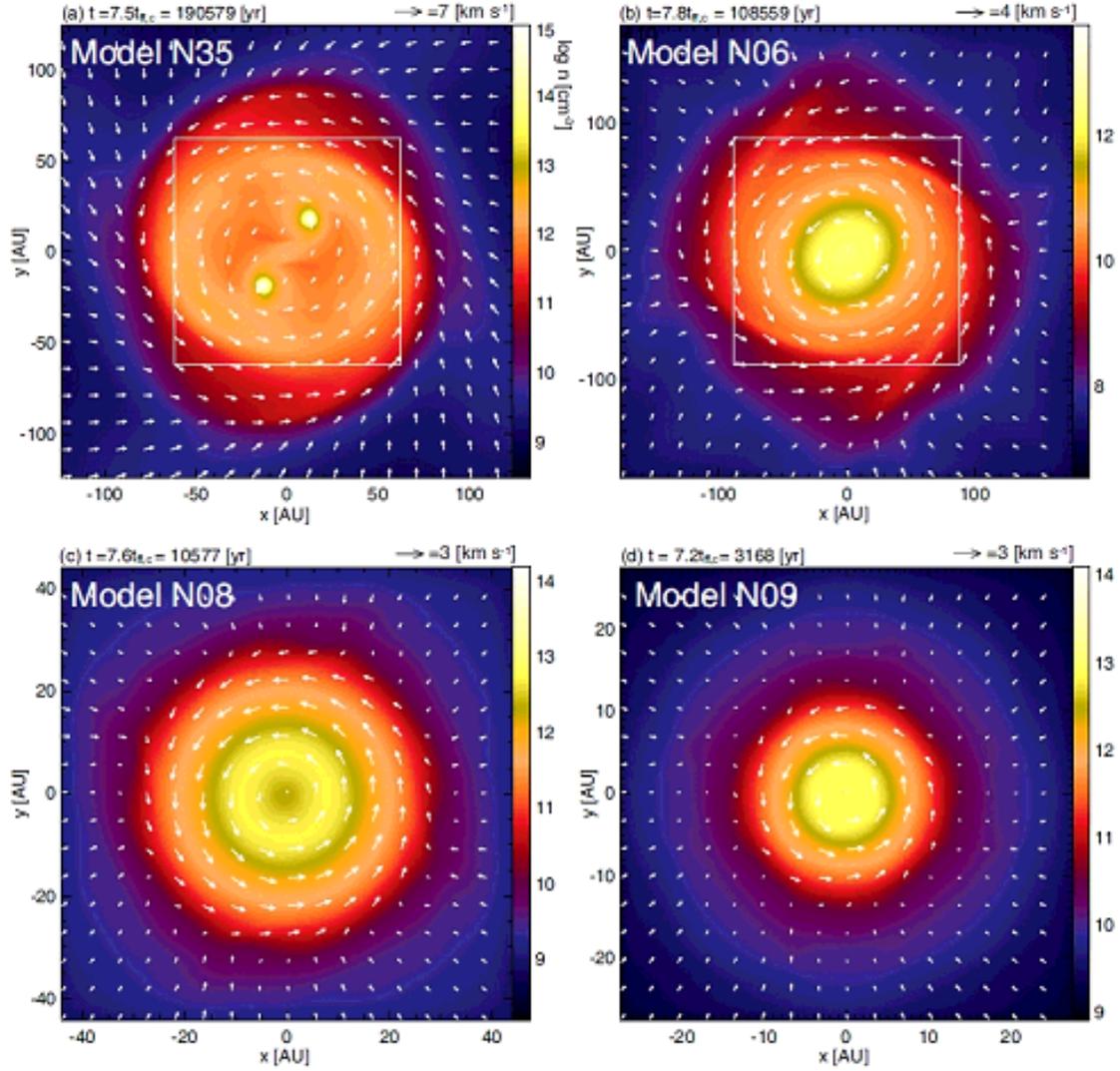}
\caption{
Density (color) and velocity (arrows) distributions on the equatorial plane for models N35, N06, N08 and N09.
The elapsed time $t$ in unit of $\tffc$ and year is plotted on the upper side of each panel.
}
\label{fig:7}
\end{figure}

Figure~\ref{fig:7} shows the density distribution on the equatorial plane around the center of the cloud at $t \sim 7\tffc$  for models N35, N06, N08 and N09.
The figure shows that a rotating disk with a size of $\sim10-100$\,AU forms around the center of the cloud by this epoch.
As seen in Figure~\ref{fig:7}{\it a}, two clumps appear in model N35, in which fragmentation occurs in the circumstellar disk about $\sim5\times10^4$\,yr after protostar formation.
In models N35 and N06, the circumstellar disk is rather massive in the early main accretion phase.
As shown in Figure~\ref{fig:6}, the circumstellar disk has a mass of $M_{\rm disk}\sim 0.4-1.0\msun$ in  model N35 and $\sim 0.4\msun$ in model N06 in the early main accretion phase for $\tilde{t} \lesssim \tffc$.
Such a massive disk becomes gravitationally unstable and tends to develop a non-axisymmetric structure and show subsequent fragmentation, as seen in Figure~\ref{fig:7}{\it a}.
Although model N06 shows no fragmentation until the end of the calculation,  non-axisymmetric (or spiral) structure develops in the circumstellar disk, as seen in Figure~\ref{fig:7}{\it b}.
The circumstellar disk mass begins to decrease after the non-axisymmetric structure develops because such structure transfers angular momentum outward, and the gas in the circumstellar disk effectively falls onto the central protostar. 
Finally, the protostellar mass dominates the circumstellar disk mass in the later main accretion phase for models N35 and N06 (Fig.~\ref{fig:6}).
For model N36, the circumstellar disk mass gradually decreases in the main accretion phase (Fig.~\ref{fig:6}), whereas it is greater than the protostellar mass at $\tilde{t}=\tffb$. 
The efficiency of angular momentum transfer for model N36 is considered to be lower than those for models N35 and N06 because only a weak spiral structure appears in this model.

On the other hand, for models N07, N37, N08 and N38, the circumstellar disk mass does not decrease greatly in the main accretion phase and is slightly greater than or comparable to the protostellar mass at the end of the main accretion phase.
%%In addition, the circumstellar disk mass for these models tends to increase in the main accretion phase, as seen in Figure~\ref{fig:6}.
Although the circumstellar disk is more massive than the protostar during the main accretion phase, no non-axisymmetric structure develops in these models.
This is because the size of the circumstellar disk is not sufficiently larger than the Jeans length, and the disk is stable against gravity (\citealt{machida10a}).
As shown in Figure~\ref{fig:7}, model N08 has an almost axisymmetric structure.
Thus, angular momentum transfer due to non-axisymmetric structure is not so effective, and a massive disk remains until the end of the main accretion phase.

As seen in Figure~\ref{fig:6}, for model N09, the protostar becomes more massive than the circumstellar disk just after the protostar formation, and the protostellar mass dominates the circumstellar disk mass by the end of the calculation.
Since model N09 has no sufficient cloud mass at the initial state, the circumstellar disk cannot increase its mass significantly by gas accretion.
For this model, the protostar and the circumstellar disk just after the protostar formation  has a mass of $M_{\rm ps}\sim0.01\msun$ and $M_{\rm disk}\sim0.01\msun$, respectively. 
Thus, the sum of protostellar and circumstellar disk masses   ($\sim0.02\msun$) is comparable to the initial cloud mass ($M_{\rm cl}=0.03$).
Therefore, mass accretion onto the circumstellar disk from the infalling envelope rapidly declines immediately after the protostar formation.
%% and the circumstellar disk cannot acquire a sufficient mass to increase its mass.
On the other hand, the protostar gradually increases its mass by the mass accretion from the circumstellar disk.
%%As a results, for this model, the protostar becomes more massive than the circumstellar disk in the early main accretion phase.

Figure~\ref{fig:6} indicates that, in each model, the protostellar mass rapidly increases for $\tilde{t} < \tffc$, and slightly increases for $\tffc \lesssim \tilde{t} \lesssim \tffb$.
For $\tilde{t} \gtrsim \tffb$, the protostellar mass rarely increases because the infalling envelope is almost depleted by this epoch.
To compare the protostellar mass evolution among models, the time averaged mass accretion rate for $\tilde{t} < \tffb$ is estimated as 
$\dot{M}_{\rm ave} = M_{\rm ps}  (\tffb) / \tffb $, where $M_{\rm ps}(\tffb)$ is the protostellar mass at $\tilde{t}=\tffb$.
For all models, the averaged mass accretion rates are in the range of $4.0\times10^{-6} < \dot{M}_{\rm ave}/(\mdot) < 8.5\times10^{-6}$.
The difference of the mass accretion rate among models is only a factor of about 2.
Thus, a protostar has a similar accretion history when the protostellar evolution is normalized by the freefall timescale of the initial cloud.
In reality, however, initially massive cloud has a longer freefall timescale and longer duration of the main accretion phase.
Thus, the protostar formed in massive cloud has a greater mass at the end of the main accretion phase.

\subsection{Mass Ejected by Protostellar Outflow}
\label{sec:out}
\begin{figure}
\includegraphics[width=150mm]{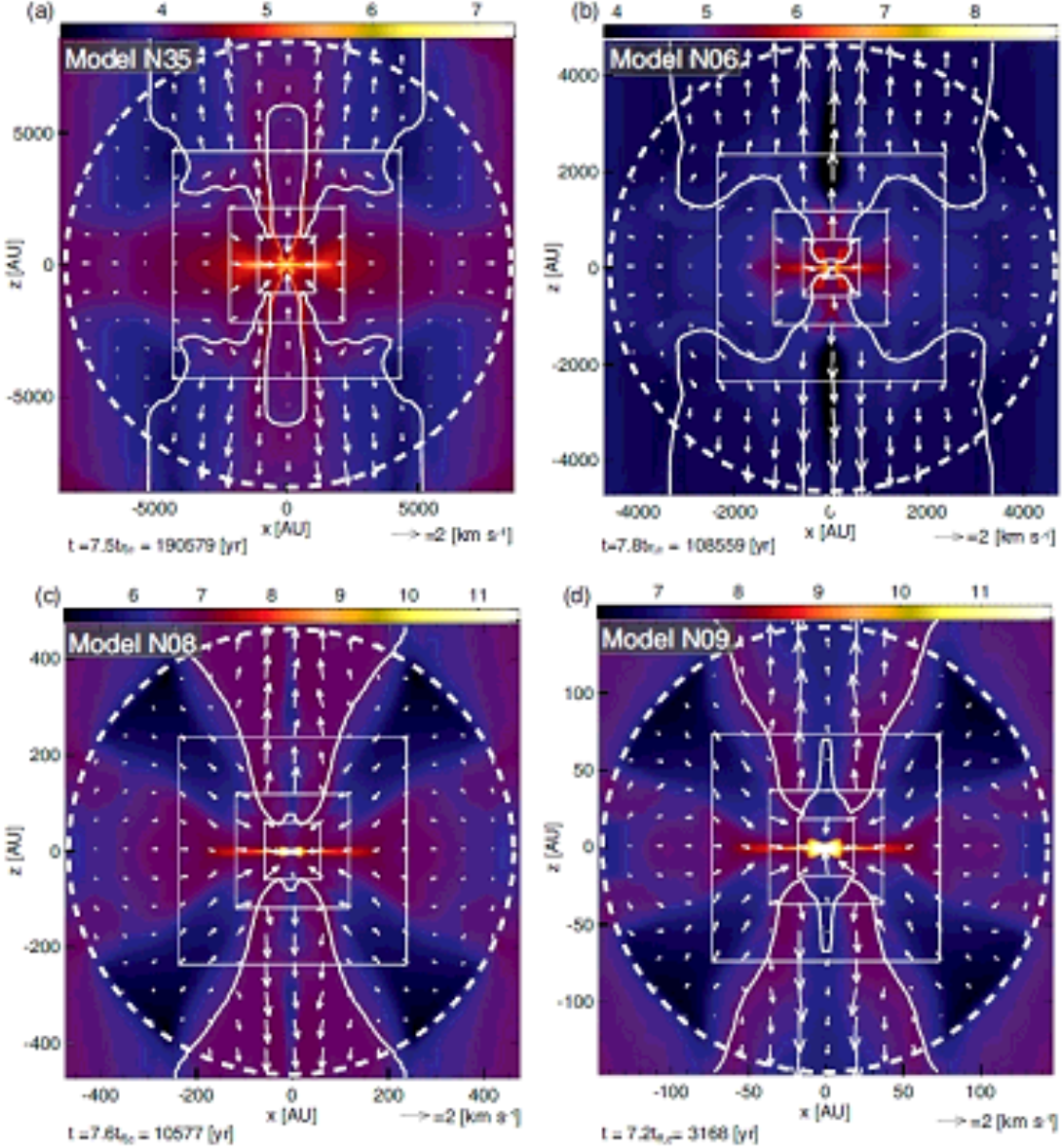}
\caption{
Density (color) and velocity (arrows) distributions on $y=0$ plane for models N35, N06, N08 and N09.
The elapsed time $t$ in unit of $\tffc$ and year is plotted on the upper side of each panel.
The white dashed circle represents initial cloud radius (i.e., BE radius).
The white solid line is the boundary of the outflow inside which the gas moves outwardly toward the center of the cloud (or the protostar) with a supersonic velocity.
The white squares in each panel denote the outer boundary of the subgrid.
}
\label{fig:8}
\end{figure}

In the main accretion phase, the mass of the protostellar outflow dominates, or is comparable to,  both the protostellar and circumstellar disk masses for models N35, N06, N36, N07, and N37, as shown in Figure~\ref{fig:6}.
This indicates that the protostellar outflow greatly affects star formation efficiency because it ejects a large fraction of the mass of the host cloud.
In other words, the protostellar outflow controls the protostellar mass or star formation efficiency in the star formation process as pointed by \citet{matzner00}.
Figure~\ref{fig:8} shows the density and velocity distribution on the $y=0$ cutting plane for models N35, N06, N08, and N09 at $t \sim7\tffc$ $\sim2\tffb$.
Note that the models and epochs in Figure~\ref{fig:8} are the same as those in Figure~\ref{fig:7}, but the spatial scales differ.
In each panel, the outflowing region is denoted by a white contour, inside of which radial velocity of the gas is supersonic ($v_r > c_{\rm s}$).

Figure~\ref{fig:8} shows that the gas is strongly ejected from the host cloud by the protostellar outflow in any model.
In addition, the outflow width (i.e., the horizontal length in the direction of travel) is comparable to the radius of the host cloud ($R_c$), and thus, the outflow has a wide opening angle.
At this epoch, on the cloud scale of $r=R_c$, the outflow has opening angles of $42\degr$ (N35), $45\degr$ (N06), $37\degr$ (N08), and $26\degr$ (N09).
Note that the opening angle continues to increase until it becomes comparable to the cloud radius.
As a result, the protostellar outflow with a wide opening angle sweeps up a large fraction of the infalling material and ejects it into the interstellar space.
In addition to this swept material, a part of the mass in the circumstellar disk is ejected directly by the outflow.

\begin{figure}
\includegraphics[width=150mm]{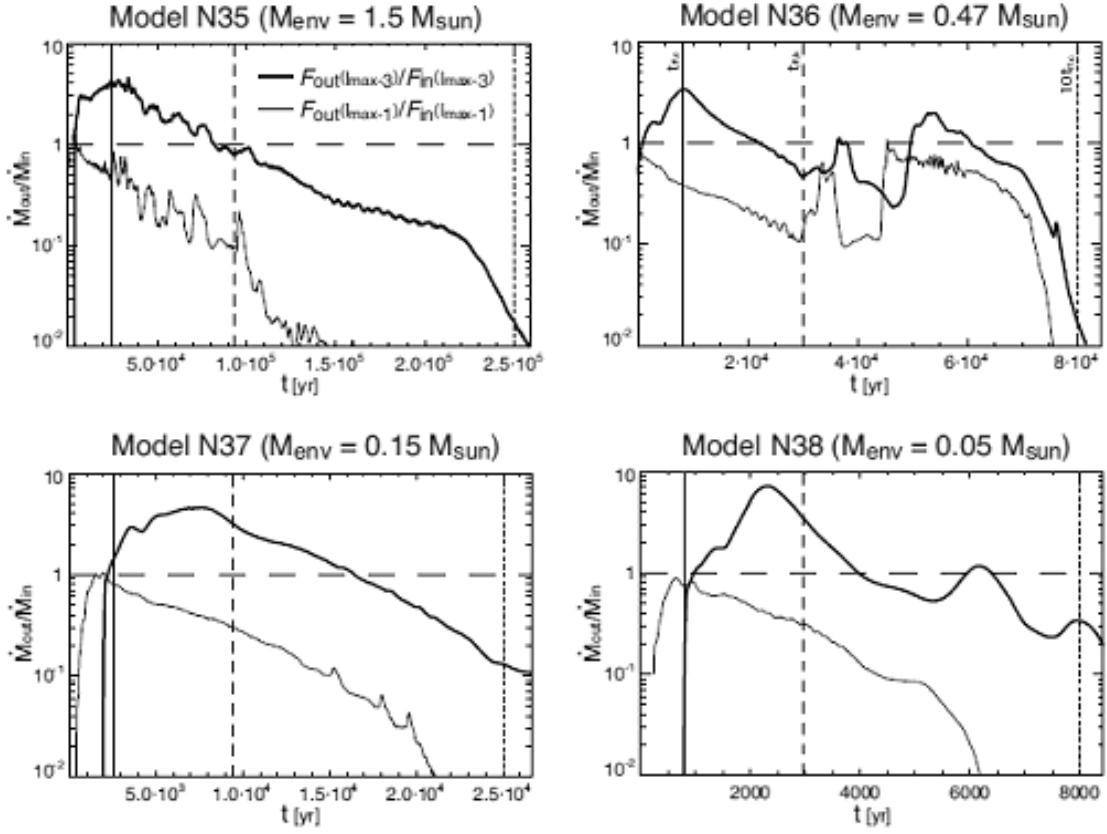}
\caption{
The inflowing and outflowing mass ratio of $l=l_{\rm max}-1$ and $l_{\rm max}-3$ grid against the time for model N35, N36, N37 and N38.
The vertical lines are the freefall timescale of the host cloud at the center ($\tffc$, solid line), cloud boundary ($\tffb$, broken line), and 10 times the freefall timescale of the host cloud at the center ($10\tffc$, dotted line).
}
\label{fig:9}
\end{figure}

Figure~\ref{fig:9} shows the ratio of the outflowing to infalling masses for models N35, N36, N37, and N38.
To investigate outflow efficiency and mass ejection rate, we calculated the outflowing/infalling mass rate on the $l=l_{\rm max}-1$ and $l_{\rm max}-3$ grid surfaces.
In each model, the $l=l_{\rm max}-1$ grid covers the entire circumstellar disk, and $l=l_{\rm max}-3$ grid has a size of $\sim8-10$ times the disk radius.
Since the protostellar outflow is originally driven by the circumstellar disk, the mass ejected from the $l=l_{\rm max}-1$ grid almost corresponds to the outflowing mass directly driven by the circumstellar disk.
On the other hand, the mass ejected from the $l=l_{\rm max}-3$ is the sum of the outflowing mass directly driven by the circumstellar disk and the swept mass by the outflow that propagates into the dense (or massive) infalling envelope.
According to \citet{tomisaka02}, we estimated the outflow $\dot{M}_{\rm out}$ and inflow $\dot{M}_{\rm in}$ masses as 
\begin{equation}
\dot{M}_{\rm out} = \int_{\rm boundary\, of \, l} \rho\, {\rm max}[\vect{v} \cdot \vect{n},0]\, ds,
\end{equation}
and
\begin{equation}
\dot{M}_{\rm in} = \int_{\rm boundary\, of \, l} \rho\, {\rm max}[\vect{v} \cdot \vect{-n},0]\, ds,
\end{equation}
respectively, where $\vect{n}$ is the unit vector outwardly normal to the surface of the $l=l_{\rm max}-1$ or $l_{\rm max}-3$ grid.
Figure~\ref{fig:9} shows that the ratio of the outflowing to inflowing mass around the circumstellar disk ($l=l_{\rm max}-1$, thin line) has a peak of $\dot{M}_{\rm out}/\dot{M}_{\rm in}\sim 1$ at $t\lesssim \tffc$.
Thus, the mass ejection rate is comparable to the mass infalling rate at this epoch.
Note that since the circumstellar disk and the outflow driving region are embedded in the grid of $l=l_{\rm max}-1$, the mass swept up by the outflow may be slightly included in the estimation of the outflowing rate.
Then,  in all the models, the ratio in the $l=l_{\rm max}-1$ grid gradually decreases with time.
The ratio decreases to $\dot{M}_{\rm out}/\dot{M}_{\rm in}\sim0.1-0.3$ at $t\sim\tffb$ and reaches $\dot{M}_{\rm out}/\dot{M}_{\rm in} < 0.01$ at $t\sim10\, \tffc$.

Figure~\ref{fig:9} also shows that the ratio of $\dot{M}_{\rm out}/\dot{M}_{\rm in}$ in the $l_{\rm lmax}-3$ grid exceeds that in the $l_{\rm max}-1$ grid for a short period after the protostar appears.
Thus, the outflowing mass (rate) increases with distance from the driving source.
This indicates that a wide-opening-angle outflow sweeps up the infalling gas and ejects it toward the center of the cloud.
Especially, for $t\gtrsim \tffc$, the rate in the $l=l_{\rm max}-3$ grid is about 10 times larger than that in the $l=l_{\rm max}-3$ grid.
This indicates that the protostellar outflow collects the matter in the infalling envelope 10 times more massive than that directly blown away from the circumstellar disk.
Note that since  the boundary of the host cloud is located more far away from $l=l_{\rm max}-3$ grid, the outflow sweeps more matter in the infalling envelope to be ejected from the host cloud.

The thick line in Figure~\ref{fig:9} indicates that the outflow on a larger scale gradually decreases for $t\gtrsim 1-3\tffc$ and disappears in $t \gtrsim 10\,\tffc$.
Thus, the lifetime of the outflow is about 10 times the freefall timescale of the center of the host cloud, or 3-5 times of the freefall timescale of the outer edge of the host cloud.
The protostar forms $\sim3-5\,\tffc$ after the cloud begins to collapse (Table~\ref{table:1}).
Thus, roughly speaking, the protostellar outflow continues to be driven by the circumstellar disk until the freefall timescale of the cloud boundary passes after the protostar formation.
This is natural that the almost all the gas inside the cloud fall onto the circumstellar disk in $\tffb$ (Figs.~\ref{fig:5} and \ref{fig:6}), and the protostellar outflow is powered by the mass accretion onto the circumstellar disk.

\subsection{Evolution of Protostellar Outflow and its Collimation}
\label{sec:out2}
\begin{figure}
\includegraphics[width=150mm]{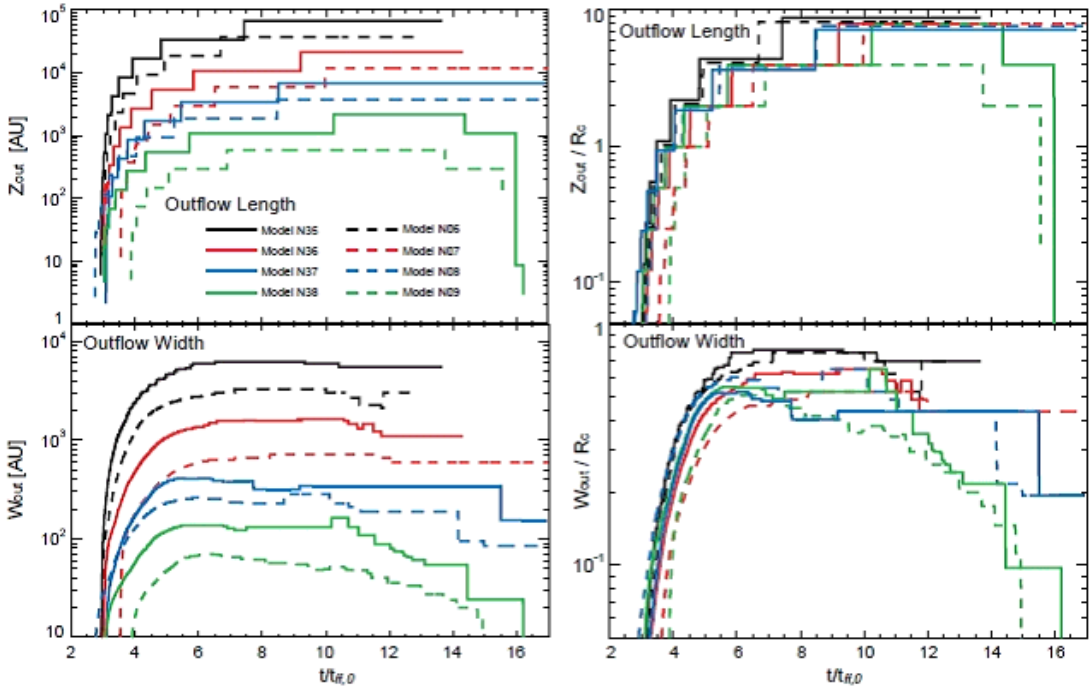}
\caption{
The evolution of the outflow length (upper panels) and width (lower panels) in unit of AU  for all models against the elapsed time normalized by the freefall timescale $\tffc$.
The outflow length and width in the right panels are normalized by the initial cloud scale $R_{\rm c}$.
}
\label{fig:10}
\end{figure}

Figure~\ref{fig:10} shows the outflow length (upper panel) and width (lower panel) against the time normalized by the freefall timescale at the center of the cloud for all models.
Note that the length and width in Figure~\ref{fig:10} right panels are normalized by the initial each cloud radius $R_{\rm c}$.
We defined the outflow as the gas having the (positive) radial velocity larger than the sound speed ($v_r > c_{\rm s}$).
The outflow expands with time and reaches $\sim 500-10^5$\,AU at $\sim10\,\tffc$ (Fig.~\ref{fig:10} upper left panel).
The final size of the outflow depends on the size (or mass) of the initial cloud.
The outflow extends up to about 10 times the initial cloud radius except for model N09 (Fig.~\ref{fig:10} upper right panel).
The outflow in a massive cloud has a longer lifetime to reach a greater distance from the protostar because a massive cloud has a longer freefall timescale and the outflow continues to be driven on $\sim1-10$ times the freefall timescale. 
The lower panels of Figure~\ref{fig:10} indicates that the outflow also expands in the horizontal direction and becomes comparable in size to the host cloud radius $R_c$.
The outflow propagates into the infalling envelope along the magnetic field lines. 
Although the magnetic field lines that drive the outflow are strongly bundled  around the circumstellar disk, they spread with the distance from the center of the cloud up to $\sim R_c$  because they are connected to the cloud scale lines.
Therefore, the outflow width also spreads in the horizontal direction and has an opening angle of $\sim45\degr$ on the host cloud scale.

\begin{figure}
\includegraphics[width=150mm]{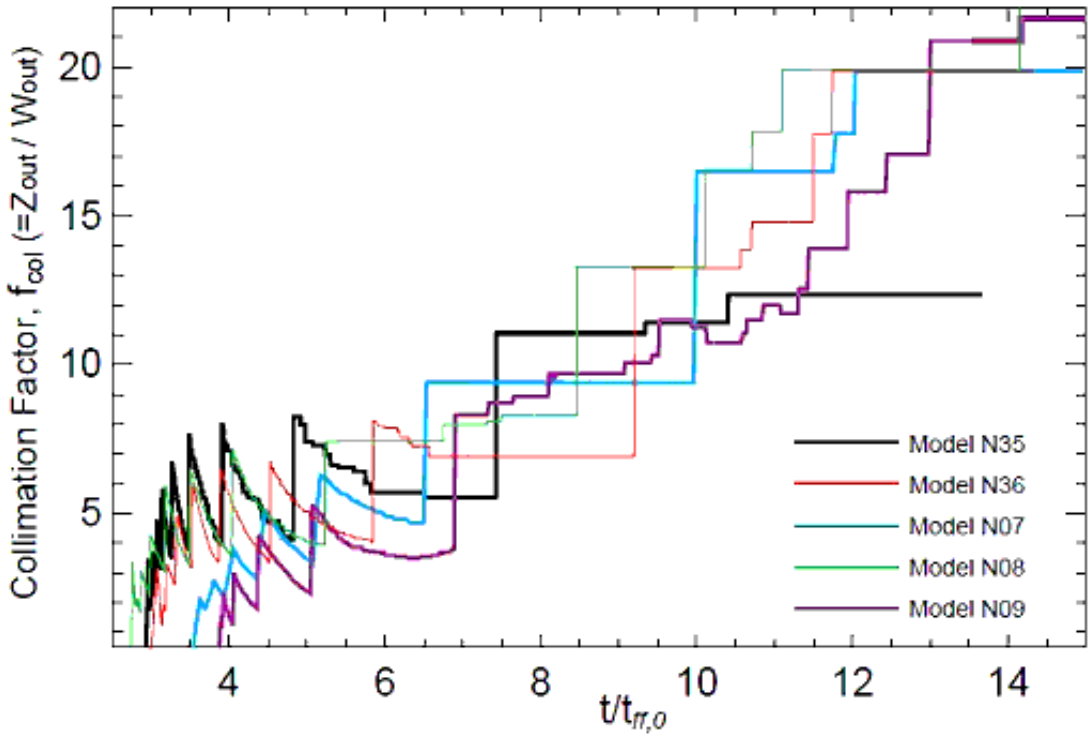}
\caption{
The evolution of the outflow collimation for models N35, N36, N07, N08 and N09 against the elapsed time normalized by the freefall timescale of the initial cloud at the center.
}
\label{fig:11}
\end{figure}

Figure~\ref{fig:11} shows the evolution of the collimation factor of the outflow, which is defined as 
\begin{equation}
f_{\rm col} = \dfrac{r_{\rm out}}{w_{\rm out}}, 
\end{equation}
where $r_{\rm out}$ and $w_{\rm out}$ are the length and width of the outflow, respectively.
The figure indicates that the collimation factors remain $f_{\rm cool}\lesssim10$ for $t<5-7\tffc$.
Then, they begin to increase for $t\gtrsim5-7\tffc$.
At the end of the calculation, the collimation factors reach  $f_{\rm cool}\sim 10-30$.
As seen in Figure~\ref{fig:2}, although the outflow has a wide-opening-angle in the early main accretion phase, it is a very well collimated in the later main accretion phase. 
In addition, the width of the protostellar outflow reflects the size of the host cloud.
Thus, we can acquire information on the prestellar cloud core from the size and width of the protostellar outflow.

\section{Discussion}
\subsection{Outflow Momentum and Comparison with Observations}
\begin{figure}
\includegraphics[width=150mm]{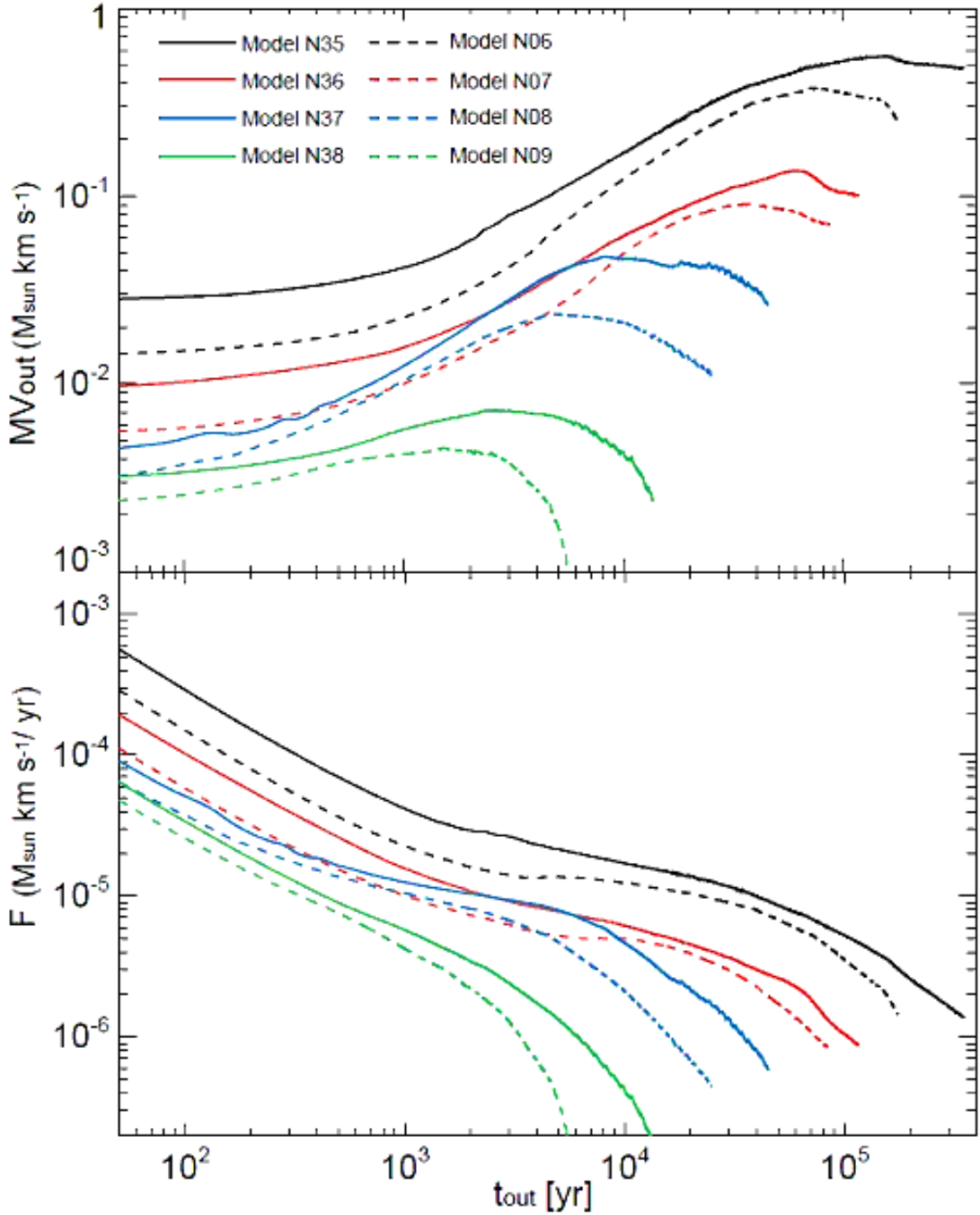}
\caption{
The momentum (upper panel) and momentum flux (lower panel) of the protostellar outflow for all models against the elapsed time after the outflow appears.
}
\label{fig:12}
\end{figure}

To investigate the evolution of the outflow strength, we estimated the outflow momentum, which is defined as
%%Figure~\ref{fig:12} upper panel shows the evolution of the outflow momentum, which is defined as 
\begin{equation}
MV_{\rm out} = \int^{v_r > c_s} \rho \, v_{\rm out} \, dv,
\end{equation}
where $v_{\rm out}$ is the outflow velocity.
Figure~\ref{fig:12} upper panel shows that the evolution of the outflow momentum against the elapsed time after the outflow appears $t_{\rm out}$, which is defined as 
\begin{equation}
t_{\rm out} \equiv t-t_{\rm out,0}, 
\end{equation}
where $t_{\rm out,0}$ is the time at the moment of the outflow appearing.
This panel indicates that an initially massive cloud has a larger outflow momentum.
The outflow momentum increases for $t\lesssim \tffb$ after the outflow appears, whereas it gradually weakens for $t\gtrsim \tffb$.
Since the main accretion phase is almost over at $t\simeq \tffb$, the outflow momentum has a peak at $t\simeq \tffb$.
At its peak, the outflow momentum is $0.003-0.6\msun$\,km\,s$^{-1}$, depending on the initial cloud mass.
The outflow originating from an initially massive cloud has a larger momentum and reaches further, while that from an initially less massive cloud has a weak momentum and disappears in a short duration.

\citet{arce07} observed many protostellar outflows in the Perseus molecular cloud complex and statistically investigated them in detail.
They showed that a large fraction of protostellar outflows have momenta in the range of $0.05 \lesssim MV_{\rm out}/(\msun {\rm km}\,{\rm s}^{-1}) \lesssim 1$.
\citet{curtis10} also investigated outflows in Perseus molecular cloud and showed that outflow momenta are distributed around $MV_{\rm out} \sim 0.1\msun\, {\rm km}{\rm s}^{-1}$.
Figure~\ref{fig:12} upper panel shows that models with  typical cloud mass of $M_{\rm cl}>0.1\msun$ (models N35, N06, N36, N07 and N37) have the outflow momentum of $MV_{\rm out} \sim 0.01-0.6\msun\, {\rm km}{\rm s}^{-1}$ for $t_{\rm out}\gtrsim 10^3$\,yr.
Thus, the observation of outflow momentum well agrees with simulation results.
Note that the observation show a snapshot of many protostellar outflows, while Figure~\ref{fig:12} upper panel shows the momentum evolution of individual outflow in each cloud.

Figure~\ref{fig:12} lower panel shows the momentum flux of the outflow, which is defined as 
\begin{equation}
F = \dfrac{MV_{\rm out}}{t_{\rm out}}.
\end{equation}
The figure shows that the protostellar outflow has a momentum flux in the range of $10^{-7} \lesssim F/(\msun {\rm km}\,{\rm s}^{-1}/{\rm yr}) \lesssim 10^{-3}$.
The momentum flux gradually decreases with time.
Roughly speaking, the outflow has a momentum flux of $F \sim10^{-4} \msun\, {\rm km}\,{\rm s}^{-1}/{\rm yr}$ in the early main accretion phase and $F  \sim10^{-6} \msun\, {\rm km}\,{\rm s}^{-1}/{\rm yr}$ in the late stage of the star formation.
\citet{bontemps96} showed that, with 45 observed outflow samples, the outflow momentum flux is typically $F \sim 10^{-4}\,\msun\,{\rm km}\,{\rm s}^{-1}$ at the early Class 0 stage and $F \sim 2\times 10^{-6}\,\msun\,{\rm km}\,{\rm s}^{-1}$ at the late Class I stage.
\citet{curtis10} also shows the similar trend of the momentum flux.
Thus, the momentum flux derived from our simulation well agrees with observations.

In the calculations, we adopted the minimum spatial scale of $\sim0.3$\,AU as described in \S\ref{sec:setting}.
Thus, we could not resolve protostar and its neighborhood.
Therefore, no high velocity jet that is driven near the protostar appears.
However, outflow momenta derived in our simulation well agree with observations.
This indicates that observed outflow momenta can be explained only by the outflow driven by the circumstellar disk without both  high velocity jet and mass entrained by the jet.
In addition, since the outflow driven by the circumstellar disk has a wide-opening-angle, it greatly contributes to determine the star formation efficiency.
%%On the other hand, since the high velocity jet has a good collimation \citep{machida08b}, it is expected that the jet rarely affects the star formation efficiency because it cannot collect a sufficient infalling matter to be ejected.
%%Note that although we calculated the cloud evolution until the initial cloud mass is depleted,  the jet may be significant in further evolution stage of the star formation.
%%It is expected that about 10\% of the accreted matter from the circumstellar disk onto the protostar is ejected by the high velocity jet.
%%However, the mass ejected by the jet rarely affects the star formation efficiency, because a large fraction of the cloud mass is already ejected by the outflow in the early stage of the star formation.

\subsection{Final Mass of Protostar, Disk and Protostellar Outflow}
\label{sec:sfe}
The masses of the protostar, circumstellar disk, and outflow at $t\simeq t_0 + \tffb$ are listed in Table~\ref{table:2}.
The table also lists the mass fraction of the protostar ($\epsilon_{\rm ps}$), protostar plus circumstellar disk ($\epsilon_{\rm disk}$), outflow ($\epsilon_{\rm out}$) and infalling envelope ($\epsilon_{\rm env}$) to that of the initial cloud at $t\simeq t_0 + \tffb$. 
As shown in Figure~\ref{fig:5}, since the mass accretion rate at this epoch ($t\simeq t_0 + \tffb$) decreases to $\dot{M}_{\rm ps} \lesssim 10^{-7}\mdot$, the protostar cannot acquire additional mass from the infalling envelope after this epoch.
Figure~\ref{fig:6} and Table~\ref{table:2} indicate that more than 90\% of the initial cloud mass is already depleted at this epoch.
Part of the cloud mass is ejected from the host cloud by the protostellar outflow; the reminder falls onto either the circumstellar disk or the protostar.
In summary, the main accretion phase is almost over by this epoch.

\begin{figure}
\includegraphics[width=150mm]{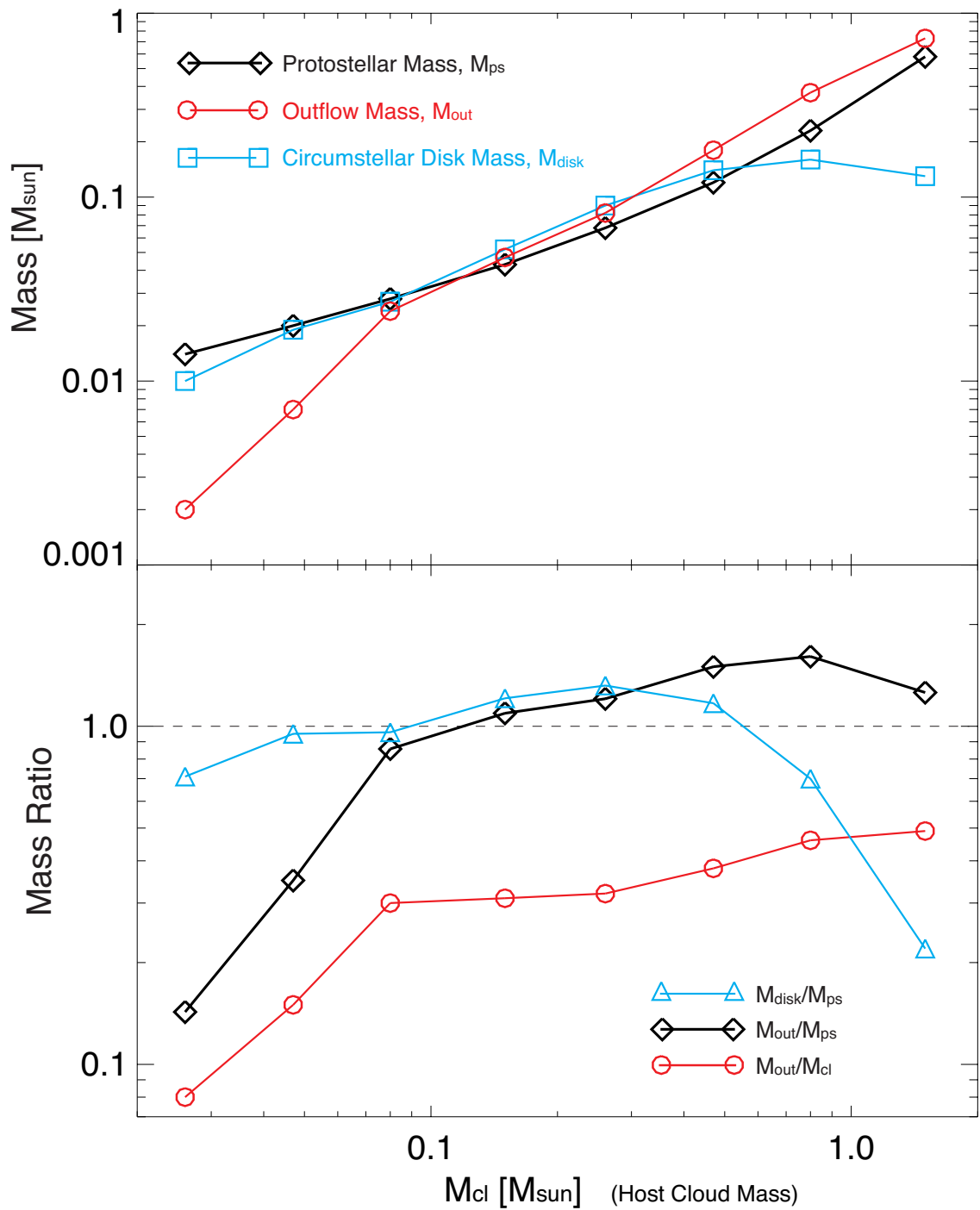}
\caption{
Upper panel: the masses of the protostar (black line), outflow (red line) and circumstellar disk (blue line) against the initial cloud mass.
Lower panel: the mass ratio of the outflow to protostar ($M_{\rm out}/M_{\rm ps}$; black line), the circumstellar disk to protostar ($M_{\rm disk}/M_{\rm ps}$, blue line), and the outflow to initial cloud ($M_{\rm out}/M_{\rm cl}$, red line) against the initial cloud mass.
%%The possible range of the star formation efficiency in each model is plotted by the vertical dashed line.
}
\label{fig:13}
\end{figure}

The upper panel of Figure~\ref{fig:13} shows the masses of the protostar, outflow, and circumstellar disk at the end of the main accretion phase ($t\simeq t_0 + \tffb$) against the initial cloud mass.
The protostar and outflow increase in mass as the initial cloud mass increases.
On the other hand, the mass of the circumstellar disk increases with the initial cloud mass when $M_{\rm cl}< 0.5\msun$, whereas it saturates when $M_{\rm disk}\sim0.1\msun$  for $M_{\rm cl}> 0.5\msun$.
As seen in Figure~\ref{fig:7}, a massive protostar formed in a massive host cloud has a massive, gravitationally unstable circumstellar disk.
Such a disk shows spiral structure and subsequent fragmentation.
The spiral structure or fragments effectively transfer angular momentum outward and promote mass accretion from the circumstellar disk onto the protostar.
Therefore, the massive circumstellar disk is self-regulated: when it is sufficiently massive, mass accretion onto the protostar is amplified, and the disk mass begins to decrease.

The upper panel of Figure~\ref{fig:13} also shows that the mass of the protostellar outflow is larger than or comparable to the protostellar mass for $M_{\rm cl}> 0.08\msun$.
This indicates that, in a massive host cloud, the protostellar outflow sweeps up a large fraction of the infalling mass and ejects it into the interstellar space.
Thus, the protostellar outflow significantly affects star formation efficiency in such a cloud.
On the other hand, in a less massive cloud, the mass of the protostellar outflow is smaller than the protostellar mass.
As described in \S\ref{sec:intro}, the first core is formed before the protostar formation.
At its formation, the first core has a mass of $\sim0.1-0.01\msun$ which is comparable to the Jeans mass at this epoch.
The first core or the circumstellar disk drives a wide-opening-angle outflow.
However, when the initial cloud mass is comparable to the mass of the first core, the infalling envelope does not contain enough mass to be swept up by the protostellar outflow.
As a result, the outflow cannot accumulate enough mass and only a small fraction of the host cloud mass is ejected.
Thus, higher star formation efficiency is realized in a less massive cloud without significant mass ejection from the host cloud.

Figure~\ref{fig:13} lower panel shows the mass ratio of protostar, circumstellar disk, and protostellar outflow at $t=t_0 + \tffb$.
As shown by the blue line, the circumstellar disk mass is comparable to the protostellar mass for $M_{\rm cl}<0.26\msun$; in this case, the circumstellar disk has an almost axisymmetric structure.
On the other hand, for $M_{\rm cl}>0.26\msun$, the mass ratio of the circumstellar disk to the protostar decreases rapidly.
The mass ratio of disk-to-protostar for model N09 is $M_{\rm disk}/M_{\rm ps}=0.71$, and that for model 35 is $M_{\rm disk}/M_{\rm ps}=0.22$.
The difference is the result of the non-axisymmetric structure and fragmentation appearing in the circumstellar disk. 
Such structure greatly promotes mass accretion from the circumstellar disk onto the protostar and decreases the circumstellar disk mass.
The cloud with $M_{\rm cl}\sim0.1-0.5\msun$ has a marginally gravitationally stable disk and has a maximum mass ratio of the circumstellar disk to the protostar.

The black line in Figure~\ref{fig:13} lower panel shows the mass ratio of the protostellar outflow to the protostar.
The ratio increases with the initial cloud mass. 
As seen in Figure~\ref{fig:9}, there are no significant difference of the mass ejection rate from the circumstellar disk by the outflow during main accretion phase among models with different initial masses; in each model, $\sim10-30$\,\% of the accreting matter is blown away from the circumstellar disk for $t\lesssim \tffb$.
However, the remaining mass of the infalling envelope and outflow driving period are different among models.
In an initially massive cloud, a large fraction of the cloud mass remains in the infalling envelope even after the protostar or circumstellar disk formation.
The outflow with a wide-opening-angle sweeps and collects the infalling material.
Thus, a large fraction of the cloud mass swept by the outflow is ejected from the host cloud.
The red line of Figure~\ref{fig:13} also indicates that a massive cloud has a large fraction of the outflowing mass. 
Therefore, the protostellar outflow effectively suppresses the star formation efficiency in a massive cloud.
As a result, the star formation efficiency in a massive cloud is lower than that in a less massive cloud.

\subsection{Star Formation Efficiency}
\begin{figure}
\includegraphics[width=150mm]{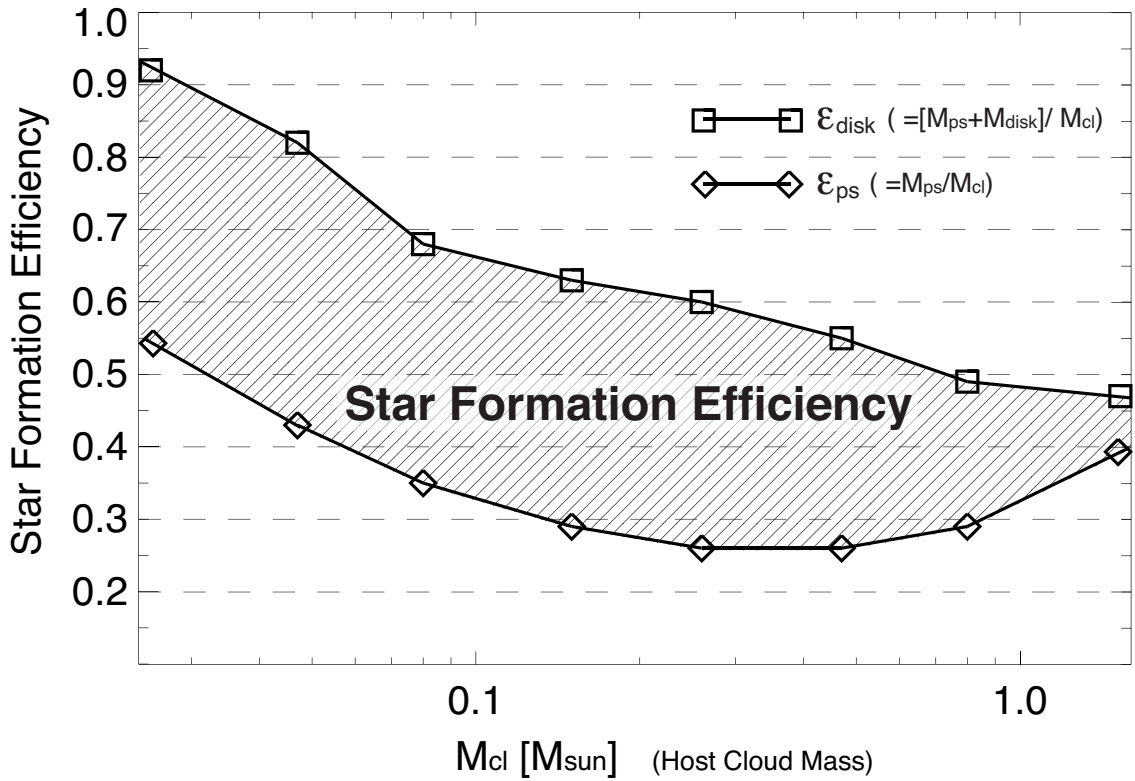}
\caption{
The star formation efficiency against the initial cloud mass.
The square symbol is the mass ratio of the protostar plus circumstellar disk to the initial cloud.
The diamond symbol is the mass ratio of the protostar to the initial cloud.
The star formation efficiency in a shadowed area is expected to be realized.
}
\label{fig:14}
\end{figure}

Figure~\ref{fig:14} shows the star formation efficiency for each model.
In the figure, the diamond and square symbols indicate the ratio of the protostellar mass to the initial cloud mass ($\epsilon_{\rm ps} \equiv M_{\rm ps}/M_{\rm cl}$; {\Large $\diamond$}) and the mass of the protostar plus circumstellar disk to the initial cloud mass ($\epsilon_{\rm disk} \equiv [M_{\rm ps}+ M_{\rm disk}]/ M_{\rm cl}$,  {\small $\square$}) at $t=t_0+\tffb$, respectively.
%% against the initial cloud mass.
At this epoch, the infalling envelope is already depleted, and gas accretion from the infalling envelope onto the circumstellar disk or protostar has already stopped. 
Thus, in a subsequent evolutionary phase of star formation (Class II and Class III phases), the protostar acquires its mass only from the circumstellar disk, not the infalling envelope.
Thus,  solid lines correspond to the lower and upper limits of star formation efficiency.
The upper limit (square symbol) is realized when the entire circumstellar disk finally falls onto the protostar, and the lower limit (diamond symbol) is realized when the entire circumstellar disk is blown away or disappears without  falling onto the protostar by any mechanism such as photo evaporation, jets around the protostar, or magnetorotational instability.
In fact, the star formation efficiency is expected to fall between the lower and upper limits.

Figure~\ref{fig:14}  shows that the upper limit of star formation efficiency decreases with the initial cloud mass.
This is because the mass ejection rate owing to the protostellar outflow increases with the host cloud mass.
As described in \S\ref{sec:intro}, the protostellar outflow is originally driven by the first core.
Thus, no outflow appears before first core formation in the collapsing cloud.
The first core has a mass of $M_{\rm fc} \simeq 0.01-0.1\msun$ \citep{saigo06} at its formation.
Thus, in a less massive host cloud, a very slight mass remains as the infalling envelope after first core formation. 
For example, when the initial cloud mass is $M_{\rm cl}=0.05\msun$ and the first core has a mass of $M_{\rm fc} = 0.04\msun$, only 20\% ($M_{\rm env}=0.01\msun$) of the initial cloud mass remains as the infalling envelope.
After first core formation, part of the gas accreted onto the first core (or the circumstellar disk) is blown away by the outflow.
Since the outflow is powered by the release of the gravitational energy of the accreting matter, no powerful outflow appears unless sufficient accreting matter exists around the driving source.
In addition, a protostellar outflow with a wide-opening-angle sweeps up the gas of the infalling envelope as it propagates into the host cloud.
However, when the infalling envelope is already depleted, the outflow sweeps up only a slight mass of the infalling envelope and ejects it into the interstellar space.
The outflow power and the amount of mass swept up by the outflow increase with the mass of the infalling envelope.
As a result, in a massive cloud, a large fraction of the initial cloud mass is ejected from the host cloud and star formation efficiency is effectively suppressed by the protostellar outflow.
As seen in Table~\ref{table:2}, for the less massive cloud model N09, 92\% of the initial cloud mass accretes onto the circumstellar disk and the protostar and only 8\% of the initial cloud mass is ejected by the protostellar outflow.
On the other hand, about half of the initial cloud  mass is ejected from the host cloud by the protostellar outflow for the massive cloud model N35.

As seen in Figure~\ref{fig:14}, the lower limit of  star formation efficiency also decreases with the initial cloud mass for $M_{\rm cl}<0.26\msun$, whereas it increases slightly for $M_{\rm cl}>0.26\msun$.
The lower limit is determined by the efficiency of mass accretion from the circumstellar disk onto the protostar.
A less massive disk appears in a less massive cloud and is stable against gravity.
In such a disk,  angular momentum is transferred by magnetic effects such as magnetic braking and protostellar outflow and the mass accretes steadily onto the protostar \citep{machida10b}.
In contrast, a massive circumstellar disk appears in a massive cloud and is unstable against gravity.
In such a massive circumstellar disk, a non-axisymmetric structure appears because of gravitational instability, as shown in Figure~\ref{fig:7}.
This structure effectively transfers angular momentum outward, promoting mass accretion from the circumstellar disk onto the protostar.
In addition, the gas accretes unsteadily onto the protostar, as described in \citet{vorobyov06} and \citet{machida10a}; these authors pointed out the possibility of episodic accretion in such massive disks.
Thus, in addition to the magnetic effects, the dynamical structure of the circumstellar disk contributes to angular momentum transfer in the massive circumstellar disk that forms in an initially massive cloud.
As a result, the accretion rate from the circumstellar disk onto the protostar increases with the initial cloud mass, and thus the protostellar mass and  star formation efficiency also increases with initial cloud mass.

Star formation efficiency in the least less massive cloud is $\epsilon = 0.54-0.92$ and that in the  most massive cloud is $\epsilon = 0.39-0.47$.
Figure~\ref{fig:14} shows that both the lower and upper limit of the star formation efficiency tend to decrease as the initial cloud mass increases.
This is because, in a massive cloud, a protostellar outflow with a wide-opening-angle can sweep up a large amount of mass and eject it into the interstellar space.
The red line in the lower panel of Figure~\ref{fig:13} shows that the mass ejection rate owing to the protostellar outflow increases with the initial cloud mass.
The mass ejection rate owing to the protostellar outflow exceeds $M_{\rm out}/M_{\rm cl} \gtrsim 0.3-0.5$ in a relatively massive cloud.
Thus, at most half of the initial cloud mass can be ejected by protostellar outflow.

Since we cannot estimate the mass ratio finally falling onto the protostar from the circumstellar disk, the realistic value of star formation efficiency is unclear.
However, our result indicates that the protostellar outflow contributes greatly to the protostellar mass and star formation efficiency in a single cloud core.

\subsection{Initial Cloud Parameters and Spatial Resolution}
In this study, we fixed the initial ratio of magnetic and rotational energies to the gravitational energy in each cloud.
As described in \S\ref{sec:sfe}, star formation efficiency depends strongly on outflow properties, which in turn depend on magnetic and rotational energies of the initial cloud.
Thus, the star formation efficiency described in \S\ref{sec:sfe} may be just a lower limit because we selected magnetic and rotational energies most suitable for driving a powerful outflow according to the previous studies \citep{tomisaka02,machida05b,machida08b}.
For example, when a very weak (negligible) magnetic field exists in the initial cloud, only a weak (or negligible) outflow appears, and higher star formation efficiency is realized without significant mass ejection from the host cloud.
Even with a slower rotation rate, a weak outflow realizes higher star formation efficiency.
In addition, the initial strength of the magnetic field affects the collimation of the outflow: outflow appearing in a weakly magnetized cloud has a relatively narrow opening angle \citep{tomisaka02}.
Thus, to investigate the relationship between star formation efficiency and protostellar outflow in more detail, we may need to investigate cloud evolution in terms of magnetic and rotational energies.
However, such calculation incurs a very high CPU cost.
In this study, we showed that the protostellar outflow can suppress star formation efficiency to $\sim25$\% ($\sim50\%$ including the circumstellar disk) at most.
This indicates that the protostellar outflow greatly affects star formation, and we need to consider the protostellar outflow in order to investigate the protostellar mass.
We will investigate cloud evolution with a large parameter space in the future.

In this study, to realize long-term evolution of the cloud until the end of the main accretion phase, we adopted sink treatment instead of resolving the protostar and the structure around it ($r\ll1$\,AU).
However, as described in \citet{tomisaka02}, \citet{banerjee06}, and \citet{machida08b}, another flow component called the high-velocity jet may appear around the protostar.
Since the high-velocity jet is well collimated \citep{machida08b}, it cannot sweep up a large amount of mass in the infalling envelope when it propagates into the host cloud.
However, about 10\% of the accreting matter is expected to be ejected by the collimated jet.
Thus, the high-velocity jet may further lower star formation efficiency during the star formation.
However, the mass ejected by the jet may rarely affect the star formation efficiency, because a large fraction of the cloud mass is already ejected by the outflow in the early stage of the star formation.
We need a considerably higher spatial resolution to include the effect of high-velocity jets to estimate the star formation efficiency in more detail.

\section{Summary}
In this study, we investigated the impact of protostellar outflow on the star formation process.
We constructed nine models with different initial cloud masses in the range of $M_{\rm cl}=0.015-1.5\msun$, and calculated the cloud evolution until the cloud mass is depleted.
In the calculation, without artificially inputting outflow (momentum) to the computational domain as seen in any other studies,  outflow is naturally or automatically driven by the circumstellar disk. 
As a result of the calculation, we found that a large fraction of the initial cloud mass is ejected from the host cloud into the interstellar space by the protostellar outflow.
Thus, the protostellar outflow significantly affects the star formation process and determination of the star formation efficiency.
The following results are obtained.

The protostellar outflow continues to be driven by the circumstellar disk for about 10 times the freefall timescale of the initial cloud after the cloud begins to collapse.
The final size of the outflow is different among models, because different models have different initial central densities and different freefall timescales. 
In each model, the protostellar  outflow reaches $\sim 500-10^5$\,AU far from the protostar in $t \sim10\,\tffc$.
When the outflow remains inside the host cloud, the outflow extends also to the vertical direction toward the direction of travel, because the outflow is anchored by large-scale (host cloud scale) magnetic field lines. 
After the outflow penetrates the host cloud and propagates into the interstellar space, it extends only in the direction of the travel keeping its width.
Thus, in this period, the outflow collimation improves to reach $\sim10-30$.
Before disappearing, the outflow has a length of 10 times the natal cloud radius and a maximum width comparable to the natal cloud radius.
Thus, we can expect  natal cloud size (or mass)  from the length and width of observed outflow.

In this study, since we did not resolve protostar itself, no high-velocity jet appears.
However, outflow momentum and momentum flux derived in our simulation well agree with observations.
This indicates that the outflow driven by the circumstellar disk is responsible for total outflow momentum, and high velocity jet rarely affects the star formation efficiency.
In addition, it is expected that the entrained mass by the high velocity jet can be ignored to estimate the mass ejection from the host cloud, because observed outflow momentum can be explained only by the disk driven outflow.
Although the high velocity jet may promote the mass ejection further, a larger fraction of the mass can be ejected only by the outflow.

The protostellar outflow can eject $\sim 10-50$\% of the host cloud mass into the interstellar space.
The mass ejection rate increases as the initial cloud mass increases; a large fraction of the initial cloud mass is ejected in a massive cloud.
This is because a massive cloud retains a large amount of the infalling matter even after the protostar formation, and  outflow can sweep and collect a large fraction of the infalling matter  when it propagates into the  infalling envelope.
Thus, the protostellar outflow can suppress the star formation efficiency to $\lesssim 50\%$.
In addition, a massive circumstellar disk comparable to the protostellar mass remains even after the infalling envelope is depleted.
Although we need to calculate further evolution of the protostellar system to determine the star formation efficiency, the star formation efficiency of $\lesssim 30$\,\% may be realized when (a part of) the circumstellar disk is blown away in further evolution stage.

\acknowledgments
We have benefited greatly from discussions with ~T. Nakano.
This work was supported by Grants-in-Aid from MEXT (20540238, 21740136).

\clearpage
%%%%%%%%%%%%%
%%% Table1%%%
%%%%%%%%%%%%%
\begin{table}
\setlength{\tabcolsep}{3pt}
\caption{Model parameters and Star Formation Epoch}
\label{table:1}
\footnotesize
\begin{center}
%%\scalebox{.5}{%
\begin{tabular}{c|cccccc|cccc} \hline
{\footnotesize Model} & 
$n_{\rm c,0}$ {\scriptsize [cm$^{-3}$]} & $R_{\rm c}$ {\scriptsize [AU]} &  $M_{\rm cl}$ {\scriptsize [$\msun$]} 
& $B_0$ {\scriptsize [G]} &  $\Omega_0$ {\scriptsize [s$^{-1}$]} &$\tffc$  {\scriptsize [yr]} & $t_0$ {\scriptsize [yr]}\\
\hline
N35  & $3 \times10^5$  & 8700 & 1.5   & $3.0\times10^{-5}$ & $1.3\times10^{-13}$ & $2.5\times10^4$ & $1.0\times10^5$ ($3.9\,\tffc$) \\
N06  & $10^6$          & 4800 & 0.8   & $5.5\times10^{-5}$ & $2.3\times10^{-13}$ & $1.4\times10^4$ & $4.5\times10^4$ ($3.2\,\tffc$) \\
N36  & $3 \times10^6$  & 2700 & 0.47  & $9.6\times10^{-5}$ & $4.0\times10^{-13}$ & $8.0\times10^3$ & $3.7\times10^4$ ($4.6\,\tffc$) \\
N07  & $10^7$          & 1500 & 0.26  & $1.7\times10^{-4}$ & $7.4\times10^{-13}$ & $4.4\times10^3$ & $1.7\times10^4$ ($3.0\,\tffc$) \\
N37  & $3 \times10^7$  & 870  & 0.15  & $3.0\times10^{-4}$ & $1.3\times10^{-12}$ & $2.5\times10^3$ & $1.0\times10^4$ ($4.0\,\tffc$) \\
N08  & $10^8$          & 480  & 0.08  & $5.5\times10^{-4}$ & $2.3\times10^{-12}$ & $1.4\times10^3$ & $4.9\times10^3$ ($3.5\,\tffc$) \\
N38  & $3 \times10^8$  & 270  & 0.047 & $9.6\times10^{-4}$ & $4.0\times10^{-12}$ & $8.0\times10^2$ & $3.1\times10^3$ ($3.9\,\tffc$) \\
N09  & $10^9$          & 150  & 0.026 & $1.7\times10^{-3}$ & $7.4\times10^{-12}$ & $4.4\times10^2$ & $2.1\times10^3$ ($4.9\,\tffc$) \\
N39  & $3 \times10^9$  & 87   & 0.015 & $3.0\times10^{-3}$ & $1.3\times10^{-11}$ & $2.5\times10^2$ & --- \\
\hline
\end{tabular}
\end{center}
\end{table}

%%%%%%%%%%%%%
%%% Table2%%%
%%%%%%%%%%%%%
\begin{table}
\setlength{\tabcolsep}{3pt}
\caption{Results}
\label{table:2}
\footnotesize
\begin{center}
%\scalebox{.5}{%
\begin{tabular}{c|cccccccccccc} \hline
Model &  $M_{\rm ps}$ {\scriptsize [$\msun$]} & $M_{\rm disk}$ {\scriptsize[$\msun$]}
 & $M_{\rm out}$ {\scriptsize[$\msun$]} &  $M_{\rm env}$ {\scriptsize[$\msun$]} &  $\epsilon_{\rm ps}$ ($\epsilon_{\rm disk}$) 
&  $\epsilon_{\rm out}$  & $M_{\rm disk}/M_{\rm ps}$ & $M_{\rm env}/M_{\rm cl}$  \\ \hline
N35 & 0.58  & 0.13   & 0.73  & 0.04   & 0.39 (0.47) & 0.49 & 0.22 & 0.03 \\
N06 & 0.23  & 0.16   & 0.37  & 0.06   & 0.29 (0.49) & 0.46 & 0.70 & 0.07 \\
N36 & 0.12  & 0.14   & 0.18  & 0.05   & 0.26 (0.54) & 0.38 & 1.17 & 0.09 \\
N07 & 0.068 & 0.09   & 0.082 & 0.026  & 0.26 (0.60) & 0.32 & 1.32 & 0.10 \\
N37 & 0.043 & 0.052  & 0.047 & 0.007  & 0.29 (0.63) & 0.31 & 1.21 & 0.05 \\
N08 & 0.028 & 0.027  & 0.024 & 0.004  & 0.35 (0.68) & 0.30 & 0.96 & 0.04 \\
N38 & 0.020 & 0.019  & 0.007 & 0.0013 & 0.43 (0.82) & 0.15 & 0.95 & 0.03 \\
N09 & 0.014 & 0.01  & 0.002 & 0.0006 & 0.54 (0.92)  & 0.08 & 0.71 & 0.02 \\
N39 & --- & --- & --- & --- & --- & --- & --- & --- \\
\hline
\end{tabular}
\end{center}
\end{table}

\clearpage
%%%%%%%%%%
% Fig. 1 %
%%%%%%%%%%

%%%%%%%%%%
% Fig. 2 %
%%%%%%%%%%

%%%%%%%%%%
% Fig. 3 %
%%%%%%%%%%

%%%%%%%%%%
% Fig. 4 %
%%%%%%%%%%

%%%%%%%%%%
% Fig. 5 %
%%%%%%%%%%

%%%%%%%%%%
% Fig. 6 %
%%%%%%%%%%

%%%%%%%%%%
% Fig. 7 %
%%%%%%%%%%

%%%%%%%%%%
% Fig. 8 %
%%%%%%%%%%

%%%%%%%%%%
% Fig. 9 %
%%%%%%%%%%

%%%%%%%%%%
% Fig.10 %
%%%%%%%%%%

%%%%%%%%%%
% Fig.11 %
%%%%%%%%%%

%%%%%%%%%%
% Fig.12 %
%%%%%%%%%%

%%%%%%%%%%
% Fig.13 %
%%%%%%%%%%

%%%%%%%%%%
% Fig.14 %
%%%%%%%%%%

\end{document}